\documentclass[usenatbib]{mn2e}

\def\lapp{\ifmmode\stackrel{<}{_{\sim}}\else$\stackrel{<}{_{\sim}}$\fi}
\def\gapp{\ifmmode\stackrel{>}{_{\sim}}\else$\stackrel{>}{_{\sim}}$\fi}

\def\deg{$^\circ$\xspace}

\usepackage{xspace}
\usepackage[usenames, dvipsnames]{color}
\usepackage{soul}
\usepackage{graphicx}
\usepackage{longtable}

\usepackage{dcolumn}
\newcolumntype{d}[1]{D{.}{.}{#1}}
\newcolumntype{X}{D{-}{-}{-1}}

\newcommand{\hdr}[1]{\multicolumn{1}{c}{#1}}

\newcommand{\fluxcal}{\texttt{fluxcal}\xspace}
\newcommand{\pac}{\texttt{pac}\xspace}
\newcommand{\psrchive}{\texttt{psrchive}\xspace}
\newcommand{\psradd}{\texttt{psradd}\xspace}
\newcommand{\coastguard}{\texttt{CoastGuard}\xspace}
\newcommand{\clean}{\texttt{clean.py}\xspace}
\newcommand{\snr}{$S/N$\xspace}

\newcommand{\dmunit}{$\mathrm{pc\,cm^{-3}}$\xspace}

\newcommand{\backend}{PSRIX\xspace}
\newcommand{\toaster}{\texttt{TOASTER}\xspace}
\newcommand{\tempotwo}{\texttt{TEMPO2}\xspace}
\newcommand{\temponest}{\texttt{TempoNest}\xspace}
\newcommand{\be}{P}

\newcommand{\rms}{\mathrm{RMS}}
\newcommand{\sect}{\S}
\newcommand{\sects}{\S\S}
\newcommand{\Msun}{\mbox{$M_\odot$}\xspace}
\newcommand{\msun}{\Msun}

\newcommand{\npsrs}{33\xspace}

\newcommand{\inprep}[1]{#1}

\newcommand{\removed}[1]{}

\title[The Effelsberg \backend Pulsar Timing Backend]
{Prospects for High-Precision Pulsar Timing with the New Effelsberg \backend Backend}

\author[P.~Lazarus~et~al.]
{P.~Lazarus$^{1}$\thanks{E-mail: plazarus@mpifr-bonn.mpg.de}, 
R.~Karuppusamy$^{1}$,
E.~Graikou$^{1}$,
R.~N.~Caballero$^{1}$,
\newauthor
D.~J.~Champion$^{1}$,
K.~J.~Lee$^{2,1}$,
J.~P.~W.~Verbiest$^{3,1}$,
M.~Kramer$^{1,4}$
\\ % The newline mark-up must be on its own line or else the tex docuemnt doesn't compile!
$^1$Max-Planck-Institut f\"ur Radioastronomie, Auf dem H\"ugel 69, 53121 
                 Bonn, Germany\\
$^2$Kavli institute for astronomy and astrophysics, Peking University, Beijing
    100871, P.R.China\\
$^3$Fakult\"at f\"ur Physik, Universit\"at Bielefeld, Postfach 100131, 33501
    Bielefeld, Germany\\ 
$^4$Jodrell Bank Centre for Astrophysics, University of
    Manchester, Manchester, M13 9PL, United Kingdom\\
} 
\begin{document}
\maketitle

%\textit{\small\input{./githash.tex}}

\begin{abstract}

The \backend backend is the primary pulsar timing instrument of
the Effelsberg 100-m radio telescope since early 2011. This new
    ROACH-based system enables bandwidths up to 500\,MHz to be recorded,
    significantly more than what was possible with its predecessor, the
    Effelsberg-Berkeley Pulsar Processor (EBPP). We review the first four years
    of \backend timing data for \npsrs pulsars collected as part of the monthly
    European Pulsar Timing Array (EPTA) observations. We describe the automated
    data analysis pipeline, \coastguard, that we developed to reduce these 
    observations. We also introduce \toaster, the EPTA timing database used to
    store timing results, processing information and observation metadata. Using
    these new tools, we measure the phase-averaged flux densities at 1.4\,GHz of all
    33 pulsars. For 7 of these pulsars, our flux density measurements are the
    first values ever reported. For the other 26 pulsars, we compare our flux
    density measurements with previously published values.  By comparing
    \backend data with EBPP data, we find an improvement of $\sim$2--5 times
    in signal-to-noise ratio achievable, which translates to an increase of
    $\sim$2--5 times in pulse time-of-arrival (TOA) precision.  We show that
    such an improvement in TOA precision will improve the sensitivity to the
    stochastic gravitational wave background.  Finally, we showcase the
    flexibility of the new \backend backend by observing several
    millisecond-period pulsars (MSPs) at 5 and 9\,GHz.  Motivated by our
    detections, we discuss the potential for complementing existing pulsar
    timing array data sets with MSP monitoring campaigns at these frequencies.

\end{abstract}


\section{Introduction}
\label{sec:Introduction}
Pulsars are extremely useful tools for studying various fields of astrophysics.
Many important results are the product of regular timing campaigns that are
used to determine models of pulsars' rotation capable of accounting for every
rotation of the star.  High-precision timing observations of millisecond-period
pulsars (MSPs) have proven to have a large number of diverse applications, such
as testing of relativistic gravity \citep[e.g.][]{ksm+06}, constraining the
equation-of-state of ultra-dense matter \citep[e.g.][]{dpr+10}, and studying
binary stellar evolution \citep[e.g.][]{fbw+11}. In general, studies of radio
pulsars have also been used to probe the interstellar medium
\citep[e.g.][]{bgr98,bm08,efk+13}. Furthermore, collections of MSPs are being
observed regularly as part of so-called pulsar timing array (PTA) projects,
which have the ultimate goal of detecting low-frequency gravitational waves,
possibly arising from the cosmic population of super-massive black-hole
binaries \citep[e.g.][]{ses13} or from cosmic strings \citep[e.g.][]{sbs12}.

To maximise the scientific potential of pulsar timing observations, high
signal-to-noise ratio (\snr) observations are required to determine pulse times
of arrival (TOAs) precisely.  Given a telescope, the \snr can be improved
either by increasing the integration time, which is limited by the total available
telescope time and the number of pulsars to observe, or by using more sensitive and/or
wider bandwidth receivers. In order to fully leverage wider bandwidths,
instruments capable of processing the increased frequency range
must be used.

The Effelsberg-Berkeley Pulsar Processor (EBPP) coherent dedispersion backend
\citep{bdz+97} has been running since 1995. Its long, uniform data sets for
some MSPs have enabled unique studies. For example, \citet{sck+13} used EBPP
data to constrain profile variations in MSPs, and thus improve limits on the
violation of local Lorentz invariance of gravity by several orders of magnitude
relative to previously published limits \citep[see][and references
therein]{wil93}. The EBPP data set has also been a key component of several
European Pulsar Timing Array (EPTA) projects, such as characterising the noise
properties of MSPs \citep{cll+15}, constraining the
low-frequency gravitational wave background \citep[GWB;][]{ltm+15}, and
searching for single sources of gravitational waves (GWs; \inprep{Babak et al.,
accepted}).

The EBPP is beginning to show its age. For instance, the EBPP bandwidth
is limited to only $\sim$64--128\,MHz, depending on the integrated Galactic
electron content along the line-of-sight to the pulsar (i.e. the pulsar's
dispersion measure, DM), whereas most current receiver systems operating in
the 1--3\,GHz band can simultaneously observe bandwidths of 200--800\,MHz
(e.g. the Greenbank Ultimate Pulsar Processing Instrument -- GUPPI --
used at the Green Bank Telescope, and its clones PUPPI and NUPPI at the Arecibo
and Nan\c{c}ay observatories, respectively \citealp{fdr10}), and in the case of the
Ultra-Broadband (UBB) receiver at Effelsberg, $\sim$2600\,MHz.  Furthermore,
the EBPP hardware is becoming increasingly unreliable, and replacement parts
are increasingly difficult to come by.

For these reasons, the EBPP backend was replaced as the main data
recorder for pulsar timing observations at Effelsberg by the \backend backend
in 2011 March. \backend is built around a Reconfigurable Open Architecture
Computing Hardware (ROACH) system, a programmable platform designed by the
Collaboration for Astronomy Signal Processing and Electronics Research
(CASPER).\footnote{https://casper.berkeley.edu/} The EBPP is still run in
parallel with \backend whenever possible.

\backend was originally designed as part of the Large European Array for
Pulsars (LEAP) project \citep{bjk+15}, which has the
objective of coherently combining signals from the five largest European radio
telescopes.\footnote{Specifically, the Lovell Telescope, the Westerbork
Synthesis Radio Telescope, the Nan\c{c}ay Telescope, the Sardinia Radio
Telescope, and Effelsberg.} To meet this goal, the primary mode of operation of
\backend is to record baseband data, however, additional modes were implemented
to record coherently dedispersed profiles folded in real-time and coherently
dedispersed single pulses.
\backend's coherent-dedispersion modes support bandwidths up to 500\,MHz and
are flexible enough to observe at different frequencies, taking advantage of
Effelsberg's many receivers. Technical details of the backend design and the
implementation its various modes of operation will be described in a future
paper.

Thanks to the increased bandwidth and more robust design of \backend compared
to the EBPP, the timing campaigns undertaken at Effelsberg using \backend are
producing data of superior quality, thus enabling even
higher-precision timing studies than previously possible. Moreover, \backend
may further improve the prospects of high-precision timing at Effelsberg by
making it is possible to conduct timing observations of MSPs at 5\,GHz and
higher, helping to mitigate noise arising from variations of the ISM along the
line of sight towards the pulsar, a serious impediment to searches for GWs with
PTAs.

The EPTA has previously incorporated the
$\sim$17-year-long EBPP data set into its timing analyses and GW searches
\citep[e.g. \inprep{Desvignes et al., submitted;}][]{jsk+08,lwj+09,lvt+11,ltm+15}.
Here we describe the \backend data and its analysis, which will be
included in future EPTA projects and be shared with the International
Pulsar Timing Array (IPTA) collaboration (\inprep{Verbiest et al., submitted}). 

In addition to the monthly observing sessions of many binary pulsars and MSPs,
several pulsars have been the target of dedicated observing campaigns with
\backend over the past four years. In particular, \backend data were included in the
IPTA effort to observe PSR~J1713+0747 continuously for 24 hours using the
largest radio telescopes around the Earth \citep{dlc+14}. Also,
    PSR~J0348+0432, a 2-\msun pulsar in a $\sim$2.5\,hr relativistic binary
    with a white-dwarf companion \citep{lbr+13,afw+13}, has been regularly
    observed for full orbits using \backend. Several full-orbit observing
campaigns of PSR~J1518+4904, a 41-ms pulsar in an 8.6-day double-neutron-star
binary, have been conducted with \backend to precisely measure the mass of the
pulsar and its companion (\inprep{Janssen et al., in prep.}).

The remainder of this paper is organised as follows.
Section~\ref{sec:observations} describes the monthly EPTA observations
undertaken with the Effelsberg telescope using \backend. The analysis of these
observations is detailed in \sect~\ref{sec:analysis}, and includes an overview
of the automated data reduction suite \coastguard, as well as the timing
database \toaster. Flux density measurements for \npsrs pulsars at 1.4\,GHz and
a comparison between \backend and the old EBPP backends are presented in
\sect~\ref{sec:results}, as are the results of observations at 5 and 9\,GHz.
The results are discussed in \sect~\ref{sec:discussion} and the paper is
finally summarised in \sect~\ref{sec:conclusion}.

\section{Observations}
\label{sec:observations}
Every month, the Effelsberg radio telescope is used to observe bright, stable
MSPs as part of the EPTA project. These observations are conducted with
\backend in its coherent-dedispersion real-time folding mode, evenly dividing
the pulse profiles into 1024 phase bins. Each session typically consists of
observations at both 1.4 and 2.6\,GHz (wavelengths of 21 and 11\,cm,
respectively). The 1.4-GHz observations use either the central feed of the
7-beam receiver (called ``P217mm'') or the single-feed 1.4\,GHz receiver
(``P200mm'').\footnote{http://www.mpifr-bonn.mpg.de/effelsberg/astronomen} Both
of these 1.4-GHz receivers are situated in the primary focus of the Effelsberg
telescope.  Only one of the 1.4-GHz receivers is installed for any given
observing session.  We use whichever receiver is available. The 2.6-GHz
observations are done with the ``S110mm'' secondary-focus receiver.  \backend
is used to record a 200-MHz band, which is divided into eight 25-MHz sub-bands.
In the case of P200mm and S110mm observations, this exceeds the available
bandwidths of 140\,MHz and 80\,MHz, respectively. See Table~\ref{tab:obssetup}
for details of the observing set-ups used. All of the receivers used in this
work have circularly polarised feeds.

Whenever possible, we record data with the EBPP coherent-dedispersion pulsar
timing backend in parallel with \backend. This allows for a more accurate
determination of the time offset between the two instruments. We have also used
these simultaneous observations to characterise the improvement of \backend
over the EBPP (see \sect~\ref{sec:ebppcompare}).

Our monthly EPTA observing sessions typically consist of 24 hours at 1.4\,GHz
and 12--24 hours at 2.6\,GHz. Each observing session includes pulsar
observations of $\sim$30\,--\,60\,min in duration. Polarisation calibration scans
are conducted prior to each pulsar observation and each consist of a 2-min
integration of the receiver noise diode offset by 0.5\deg from the pulsar
position. The diode is pulsed with a 1-s repetition rate and a 50\% duty cycle.

Since 2013, at 1.4\,GHz, we also performed on- and off-source scans of a
radio source with a stable, well-known flux density, usually 3C~218 (i.e.
Hydra~A). These flux calibration observations use the noise diode as
described above.

Every month, we observe $\sim$45 pulsars at 1.4\,GHz and $\sim$20
pulsars at 2.6\,GHz. Pulsars that are never, or rarely, detected at 2.6\,GHz
during a 6--12 month probationary period are dropped from the regular
observing schedule. Here we focus on the data sets of \npsrs MSPs and
binary pulsars acquired between 2011 and 2015. Tables~\ref{tab:obssumm 21cm}
and~\ref{tab:obssumm 11cm} show a summary of our 1.4~and 2.6\,GHz observations
of these pulsars, respectively.

As we will discuss in \sect~\ref{sec:highfreqpta}, since ISM effects weaken
with increasing radio frequency, high-frequency observations of pulsars may be
extremely useful to avoid and mitigate the effects of variability in the
interstellar medium (ISM), which limit the sensitivity of attempts at detecting
GWs with pulsars.  In 2015 January, we conducted observations of 12 MSPs at
5\,GHz (6\,cm) using the ``S60mm'' secondary-focus receiver with the aim of
assessing their utility to the PTAs. We used \backend in its 500-MHz
coherent-dedispersion real-time folding mode for these 5 observations.  We also
observed four of these pulsars at 9\,GHz (3.6\,cm) with the ``S36mm'', also a
secondary-focus receiver, again with 500\,MHz of bandwidth. Our high-frequency
observations are listed in Table~\ref{tab:highfreq}. We selected the pulsars
for these exploratory high-frequency observations based on their 1.4 and
2.6\,GHz detection significances, which we scaled to higher frequencies using
the radiometer equation, the receiver performance, and published spectral
indices. \footnote{For pulsars without spectral indices available in the
literature, we used a spectral index of $\alpha = -2$ for our estimates.}
Specifically, we required an estimated $\snr\;\gapp\!10$ for a 30-min observation
when selecting pulsars for the preliminary 5 and 9-GHz observations reported
here. High-frequency observations of other (fainter) MSPs are being conducted
and will be reported elsewhere.

\section{Data Analysis}
\label{sec:analysis}

\subsection{CoastGuard: An Automated Timing Data Reduction Pipeline}
\label{sec:coastguard}
We developed an automated pipeline, \coastguard,\footnote{Available at
https://github.com/plazar/coast\_guard} to reduce \backend data. \coastguard
is written in python and is largely built around programs from the \psrchive
package\footnote{http://psrchive.sourceforge.net/} \citep{hvm04}, using
its python wrappers to read \backend data files, which are
\psrchive-compatible. \coastguard contains components that are sufficiently
general for use with \psrchive-compatible data files from other observing
systems despite that it was primarily designed for Effelsberg \backend data. In
particular, the radio frequency interference (RFI) removal algorithm described
below has been applied to data from the Parkes Telescope \citep{cbb+14} and has
also been adopted by the LOFAR pulsar timing data reduction pipeline
\citep{kvh+15}.

\coastguard contains considerable error checking, logging, logistics, and
control logic required to automate large portions of the pipeline, which is
marshalled by a control script and a MySQL database.

In its coherent-dedispersion real-time folding mode the \backend backend writes
data files every 10\,s for each 25-MHz sub-band.  These fragments are then
grouped together and combined using \psradd. At this stage, the data are
re-aligned using an up-to-date pulsar ephemeris, if necessary, and 12.5\,\%
of the channels at the edge of each sub-band are zero-weighted to reduce the
effect of aliasing. 

Next, the metadata stored in these consolidated files are cross-checked
against telescope observing logs and all discrepancies are corrected. This is
primarily to repair issues with the observation metadata that were common
during the commissioning of \backend. These issues have since been resolved.

The data files are then cleaned of RFI. In the pipeline, our cleaning process
excludes RFI by setting the weights of individual profiles to zero. That is,
the data from RFI-affected sub-integration/channel combinations are ignored in
the rest of the analysis without altering the data values.  Therefore, it is
possible to reverse the automated RFI masking.

\coastguard's RFI-excision script, \clean, includes four distinct algorithms
that can be chained together to clean corrupted data.  Each algorithm has
several parameters that can be used to optimise its performance.  In our
automated data analysis, we use two of the four available cleaning algorithms,
namely \texttt{rcvrstd} and \texttt{surgical}.  The other two algorithms,
\texttt{bandwagon} and \texttt{hotbins}, are occasionally applied manually to
observations requiring special attention.  Our standard RFI excision algorithm
proceeds as follows: 

First, \clean's \texttt{rcvrstd} algorithm is used to zero-weight frequency
channels beyond the receiver response and channels falling within a list of
receiver-dependent bad frequency intervals.

Second, the \texttt{surgical} algorithm is used to find profiles corrupted by
RFI in the folded data cube. To avoid being biased by the presence of the
pulsar signal, the amplitude and phase of the integrated pulse profile is fit
using a least-squares algorithm to individual profiles containing a significant
detection and the difference is computed.  It is these pulsar-free residuals
that are treated in the remainder of the algorithm.  Next, RFI-contaminated
data are identified with a set of four metrics, which are computed for each
sub-integration/channel pair (i.e.  each total-intensity profile stored in the
data file). These metrics are: 1) the standard deviation, 2) the mean, 3) the
range, and 4) the maximum amplitude of the Fourier transform of the
mean-subtracted residuals. These four metrics were selected due to their
sensitivity to the RFI signals present in Effelsberg data, which include, but
are not limited to: excess noise, occasional data drop-outs, and infrequently,
rapid (sub-ms) periodic bursts. For each metric, a $N_\mathrm{sub} \times
N_\mathrm{chan}$-sized matrix of values is produced.  Trends in the rows and
columns of these matrices are removed by subtracting piece-wise quadratic
functions that were fit to the data. The subtraction of these trends account
for slow variations in time, as well as the shape of the bandpass. These
rescaled matrices are then searched for outliers, which are defined as being
$>$5\,$\sigma$ from the median of either their sub-integration or channel.
Finally, profiles that are identified as an outlier by at least two of the four
metrics are zero-weighted.

The \texttt{bandwagon} algorithm completely removes sub-integrations and
channels that already have a sufficiently large fraction of data masked, and
the \texttt{hotbins} algorithms replaces outlier off-pulse profile phase-bins
with locally sourced noise.\footnote{Because the \texttt{hotbins} algorithm
replaces data, it is irreversible.} Neither of these two algorithms are part of our
standard automated data reduction.

Once the observations are cleaned, they are reviewed before proceeding with
the rest of the automated analysis. This is to identify observations that still need
to be cleaned manually. In practice, only a small fraction of observations
require additional RFI zapping. This quality-control stage also provides an
opportunity to identify observations where the pulsar is not detected or where
the data are contaminated by RFI beyond repair. Observations falling into these
two categories do not continue further in the data reduction process. 

The above data reduction process (combine, correct, clean, quality control) is
also applied to polarisation calibration scans of the noise diode.
The cleaned and vetted calibration scans are fully time-integrated, and then 
loaded into the appropriate \psrchive \pac-compatible ``database'' files.
The pipeline maintains one calibration database file for each pulsar.

Polarisation calibration of the cleaned pulsar data files is performed with
\psrchive's \pac
program,\footnote{http://psrchive.sourceforge.net/manuals/pac/} using its
``SingleAxis'' algorithm, which appropriately adjusts the relative gain and
phase difference of the two polarisation channels by applying the technique
of \citet{bri00}.  These calibrated observations are manually
reviewed a second time to verify that no artifacts have been introduced. 

Flux calibration has not been incorporated into the automated data analysis
pipeline. Nevertheless, we have manually performed flux calibration wherever
possible. In our analysis we used \psrchive's \fluxcal and \pac programs.
\fluxcal compares the power levels of on- and off-source observations of a
standard candle target to determine the system equivalent flux density over the
observing band. This information is used to determine the flux density scale of
the polarisation-calibrated pulsar observations.

For this paper we refolded all data with up-to-date ephemerides.  

%Information about AGN's spectral indexes and reference flux measurements that
%are stored inside \psrchive are cross-checked with the NED catalogue reported
%data.  

\subsection{TOASTER: The TOAs Tracker Database}
\label{sec:toaster}
We have developed a python package, TOAS TrackER
(\toaster),\footnote{\toaster and its documentation are publicly available at
https://github.com/plazar/toaster} for computing and storing TOAs in a fully
described and reproducible way. At its core, \toaster consists of an SQL
database and full-featured python toolkit for reliably interacting with the
data and database.  

Beyond simply storing TOA information, \toaster's database also records
information about telescopes and observing systems, observation information
(e.g. frequency, epoch, integration time, the ephemeris used for folding), the
standard profile used to determine each TOA, as well as version
numbers of relevant software, such as \psrchive and \tempotwo \citep{hem06}.  

Once the database is populated, \toaster can also launch TOA generation
processes that use a variety of ``manipulators'' to prepare the data prior to
automatically computing TOAs using standard \psrchive tools. The most basic
manipulator fully integrates data in frequency and time. However, more
sophisticated manipulations can be included to adjust the data according to an
updated, possibly time-varying DM, integrate a fixed number of pulses or
variable number of pulses depending on the resulting \snr. Manipulators can
also be used to scale the measured profiles to have uniform off-pulse variance,
as was done by \citep{abb+15}. Typically these types of manipulations are included
in the data reduction pipelines that prepare observations prior to determining TOAs.
By performing these data reduction steps in \toaster, the details of the
manipulations performed on the data and the resulting TOAs are logged in the
database, making it easy to store the resulting TOAs, as well as systematically
compare the effect of different manipulations on the eventual timing analysis.
Furthermore, the \toaster database includes a reference to the template used to
compute each TOAs. The end result is a completely described and reproducible
TOA-generation procedure. This makes \toaster a useful tool for high-precision
timing projects like the EPTA and IPTA that are constantly adding new data, as
well as developing new data reduction algorithms.

The \toaster toolkit scripts can be used to easily query the information stored
in its database.  For example, \toaster provides scripts to list and summarise
the TOAs in the database. These scripts can also be used to generate TOA files in
multiple formats, including a \tempotwo format that includes all the
annotations (``TOA flags'') requested by the IPTA. 

\toaster can be used to load TOAs directly into the database (i.e.  without
information concerning the observations, templates, etc.). This feature is
useful for including previously computed, and finalised, data sets, such as the
EPTA legacy TOAs \inprep{(Desvignes et al., submitted)}.

We set up the \toaster software and database to manage the reduced (i.e.
cleaned and calibrated) \backend data, which are automatically loaded into the
\toaster database by the data reduction pipeline described in
\sect~\ref{sec:coastguard}. 

%\toaster interfaces with the database using the \texttt{SQLAlchemy}
%python library, which supports many common implementations of the SQL database
%language without requiring any changes to the code. Thus, \toaster is suitable
%for large projects such as the \backend data analysis results described here
%and PTA projects such as the EPTA and the IPTA (using a MySQL or MSSQL
%database), as well as for smaller projects that might only require detailed
%book keeping for a single pulse (using a SQLite database file).

\section{Results}
\label{sec:results}

Over the past four years, we have collected timing data on 45 pulsars at
1.4~and 2.6\,GHz using the \backend backend with the 100-m Effelsberg
radio telescope. Here we report on a selection of \npsrs pulsars. Most of these
pulsars have been monitored monthly in both bands for the entire 4-year period.
An overview of our 1.4~and 2.6\,GHz observations can be found in
Tables~\ref{tab:obssumm 21cm} and~\ref{tab:obssumm 11cm}, respectively.

\subsection{Flux Density Measurements}
\label{sec:fluxes}

We measured flux densities for all \npsrs pulsars at 1.4\,GHz. For each pulsar,
we report the mean flux density, $\left < S \right >$, and the median flux
density, $S_\mathrm{med.}$, to account for observed modulation due to
interstellar scintillation.  We estimate the precision of the mean flux
densities as the standard error on the mean, that is,

\begin{equation}
    \delta \left < S \right > = \sigma_S/\sqrt{N_\mathrm{cal.}} ,
    \label{eq:fluxuncertainty}
\end{equation}

\noindent where $\sigma_S$ is the standard deviations of the individual flux
measurements and $N_\mathrm{cal.}$ is the number of calibrated observations.

\begin{figure}
    \includegraphics[width=\columnwidth]{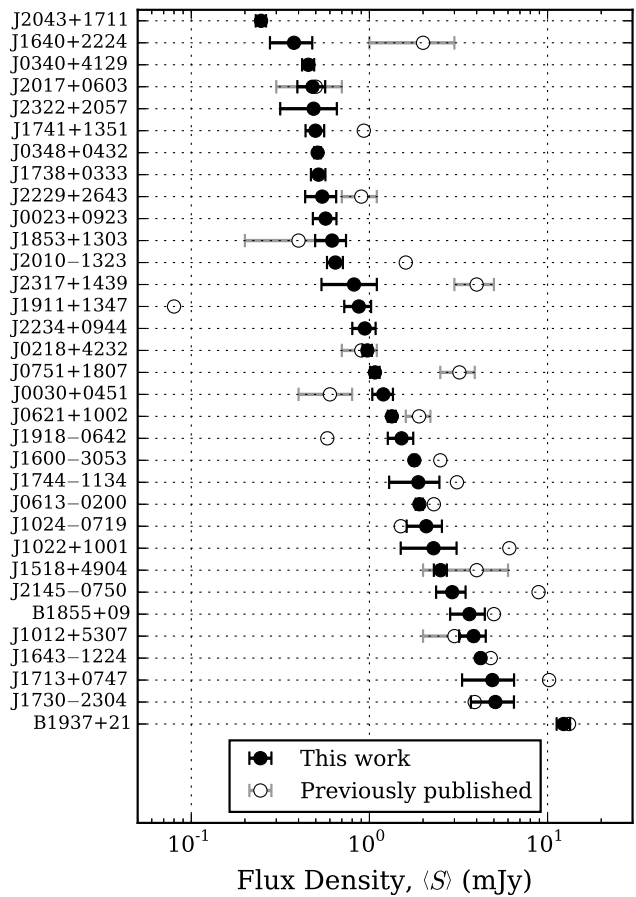}
    \caption{
        Measured 1.4\,GHz flux densities of \npsrs pulsars estimated from
        averaging over multiple \backend observations (filled circles). The
        uncertainties are estimated as the standard error on the mean (see
        Eq.~\ref{eq:fluxuncertainty}).  Additional details of the calibration
        process can be found in \sects~\ref{sec:analysis} and~\ref{sec:fluxes}.
        The previously published flux densities (unfilled circles) do not all
        have properly measured uncertainties. See Table~\ref{tab:obssumm 21cm}
        for notes and references.
    \label{fig:fluxes}}
\end{figure}

The flux densities we measure are reported in Table~\ref{tab:obssumm 21cm},
along with previously measured values at 1.4\,GHz.  Seven of the pulsars we
report flux densities for do not have previously published measurements, and
three other pulsars have previously published measurements that were not
calibrated against observations of standard candle sources. Most of the rest
of our flux density measurements are consistent with previously reported values
(see Fig.~\ref{fig:fluxes}).  Inconsistencies may arise from scintillation,
which impacts both the observed flux density as well as the apparent
uncertainty.  The effect of scintillation is most apparent when only a small
number of observations are used to estimate the flux density and is further
exacerbated when observations make use of short integrations and/or small
bandwidths.

Data from 2013~Nov. to 2014~Aug. could not be calibrated due to
saturation and/or non-linearities in the data resulting from insufficient
attenuation of the telescope signal.  Fortunately, this was only an issue
when observing extremely strong sources (e.g. flux calibrators with the noise
diode). We find no anomalies in the observed pulse profiles, allowing these
observation from late-2013 to mid-2014 to be used for timing.

\subsection{Clock Stability}
\label{sec:stability}
The \backend system suffered four clock offsets over its first four years
of operation. The first offset occurred between 2012~Oct.~27 and Nov.~10, and
was due to switching clock sources without measuring the phase difference
between their signals. The second offset, which occurred on 2013~July~27, was
caused by cutting the power to the clock signal generator and not re-syncing
the phase of the signal after the system was restarted. The third offset was
deliberately introduced on 2014~Mar.~4 when the clock signal was synchronised
to the original clock phase.  Finally, the fourth offset on 2014~Nov.~20 was
also deliberately introduced by installing a new clock signal generator. 

The first two clock offsets were initially measured by fitting timing data for the
orbital phase of PSR~J0348+0432. These measurements were sufficiently precise
to determine the offsets to within one phase rotation of PSR~J0348+0432 ($P \simeq
39$\,ms), allowing the values to be further refined by fitting arbitrary time
offsets (``JUMPs'') to the timing residuals
of PSR~J0348+0432 and then with PSR~J1744$-$1134 ($P \simeq 4.1$\,ms). The
final values of the clock offsets have been measured by fitting JUMPs
individually to the timing data of four pulsars, namely PSRs~J0613$-$0200,
J1643$-$1224, J1713+0747, and J1744$-$1134. These were selected on the basis of
being of the most precisely timed pulsars in the \backend data set. The JUMPs
were fit simultaneously with pulsar parameters and noise models using
\temponest \citep{lah+14}. The resulting JUMP values, all of which were
measured relative to the original clock signal, were averaged together
resulting in measurements of $\Delta T_A=97.2851(6)$\,ms and $\Delta
T_B=409.2691(8)$\,ms.  These measurements have been confirmed with data from
LEAP by measuring and comparing the phase delays between the signals of
simultaneous observations with several European radio telescopes before and
after the epochs of the \backend clock offsets \citep[see][ for an overview of
the project]{bjk+15}.\footnote{The precision of the LEAP-based measurements is
expected to surpass what is possible with timing-based JUMP measurements.
However, the uncertainties of the LEAP-based measurements are not yet well
determined, so here we report the values and uncertainties derived from the more
standard and conservative JUMP measurements.} The third and fourth offsets were
directly measured at the telescope by comparing clock signals with an
oscilloscope. The results are high-precision measurements of $\Delta
T_C=0.000612(1)$\,ms and $\Delta T_D=0.000127(1)$\,ms, which are consistent
with offset values derived from fitting JUMPs to pulsar timing data.

\begin{figure}
    \includegraphics[width=\columnwidth]{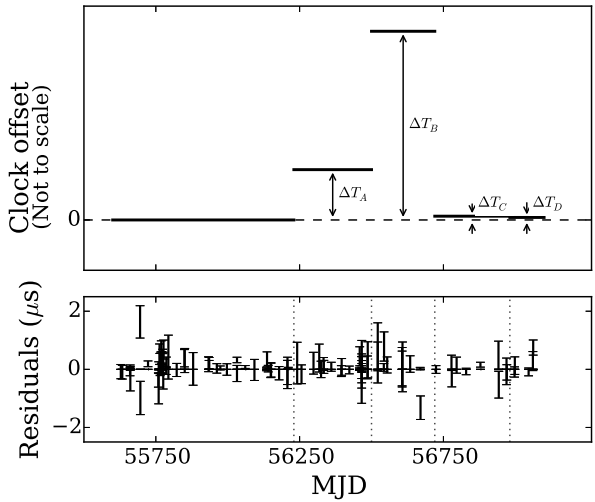}
    \caption{\textit{Top} -- A schematic of the four clock offsets suffered by
        the \backend system. The offset values are: $\Delta
        T_A=97.2851(6)$\,ms, $\Delta T_B=409.2691(8)$\,ms, $\Delta
        T_C=0.000612(1)$\,ms, and $\Delta T_D=0.000127(1)$\,ms. Offsets
        \textit{A} and \textit{B} are larger than the spin periods of the
        pulsars reported here ($P\sim$1\,--\,50\,ms), thus resulting in phase
        ambiguities and different apparent offsets in residuals for different
        pulsars.  See \sect~\ref{sec:stability} for the origins of the clock
        offsets and how their magnitudes were determined.
        \protect\\ 
        \textit{Bottom} -- Timing residuals from \backend observations
        at 1.4\,GHz of PSR~J1713+0747 after accounting for the clock offsets,
        showing that no significant offsets remain.
    \label{fig:clockoffset}}
\end{figure}

A schematic of the \backend clock offsets is shown in
Fig.~\ref{fig:clockoffset}. The timing residuals of PSR~J1713+0747 after the
JUMPs are removed show no evidence of the clock offsets, as shown in the bottom
panel of Fig.~\ref{fig:clockoffset}. Similarly, the residuals of all other
pulsars are also free of the effect of the clock offsets after applying the
offsets listed above.

Note that the EBPP uses an independent reference clock, and thus was not
affected by any of the four offsets seen in the \backend data.

\subsection{Comparison With the EBPP Backend}
\label{sec:ebppcompare}
\begin{figure}
    \includegraphics[width=\columnwidth]{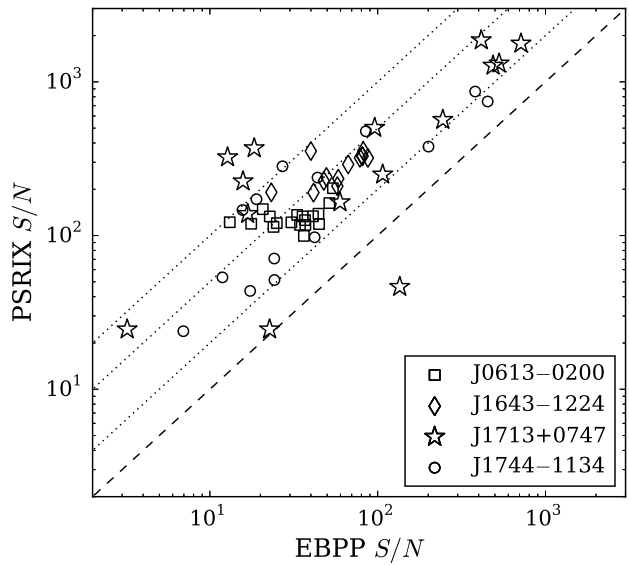}
    \caption{Comparison of \snr for simultaneous observations with the \backend and
             EBPP backends. Equal \snr is shown with the dashed line. The
             dotted lines represent 2$\times$, 5$\times$, and 10$\times$
             improvements. The one observation of PSR~J1713+0747 where the \snr
             is larger in the EBPP observation is due to there being a single
             scintillation maximum within the 200-MHz \backend band that falls
             inside the smaller EBPP observing window. Note that we did not
             weight the frequency channels by \snr when integrating the band.
             \label{fig:snrcompare}}
\end{figure}

We have compared the \snr %and TOA uncertainty 
derived from data recorded simultaneously with \backend and the old EBPP
backend (see Fig~\ref{fig:snrcompare}).
%~and~\ref{fig:toacompare}).
In particular, we used multiple observations of four MSPs that are best
and most frequently timed with \backend, PSRs~J0613$-$0200, J1643$-$1224,
J1713+0747, and J1744$-$1134.  We have found that \backend provides
significantly stronger detections, roughly 2--5 times higher \snr, than the
simultaneously recorded EBPP data. A similar comparison of TOA uncertainties
derived for simultaneous \backend and EBPP data also shows improvements of a
factor of 2--5.

\begin{figure}
    \includegraphics[width=\columnwidth]{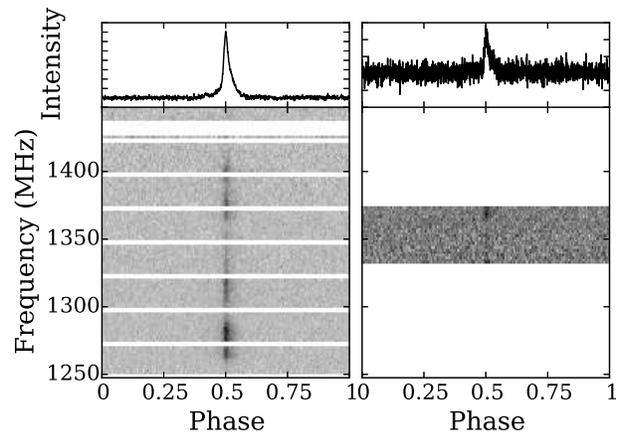}
    \caption[Comparison between simultaneous observations with the \backend and EBPP backends]
            {\textit{Left} -- The integrated pulse profile and frequency vs.
                phase plot for a 28-min observation with \backend of
                PSR~J1713+0747 on 2013 Jan 6. This detection has a \snr~=~225
                thanks to its 200\,MHz bandwidth. The frequency channels
                missing are removed due to interference and roll-off at the
                edges of the sub-bands (see \sect \ref{sec:analysis}).
                \protect\\ 
             \textit{Right} -- The integrated pulse profile and frequency vs.
             phase plot for the same observation of PSR~J1713+0747 using data
             from EBPP, which was recording in parallel. The EBPP provides a
             significantly weaker detection with \snr~=~20, owing to its limited
             $\sim$40\,MHz bandwidth.
             \label{fig:bandcompare}}
\end{figure}

There are several reasons why \backend outperforms the EBPP: 

\noindent 1) The 200-MHz bandwidth of \backend is considerably larger than the
EBPP's usable bandwidth ($\sim$40--50\,MHz for most pulsars, and
$\sim$95\,MHz for pulsars with $\mathrm{DM} \lapp 10$\,\dmunit).
A comparison of the observing bands from both backends is shown in
Fig.~\ref{fig:bandcompare}. \backend's larger bandwidth allows more signal to
be integrated, reducing radiometer noise, and also increases the chance of
observing constructive scintels. 

\noindent 2) The \backend data are recorded with 8 bits, making them even more
resilient in the presence of strong RFI than the EBPP with its 4-bit data.

\noindent 3) The 10-s sub-integrations of \backend are much shorter than the
2-min sub-integrations of the EBPP. Thus, the expense of removing impulsive
RFI is diminished. Also, the shorter sub-integrations make re-aligning the
pulse profiles with an updated timing model more accurate. 

\noindent 4) \backend is a more robust instrument than the EBPP. This is
especially true now that the latter is nearly 20 years old, and hardware and
networking issues occasionally preclude it of recording data. In these
instances, data files are cut short, or not written at all. 

The increase in bandwidth of \backend over the EBPP is even more apparent when
full polarisation information is recorded. Polarisation observations with the
EBPP are limited to only 28\,MHz, whereas with \backend full polarisation
information can be recorded for up to 500\,MHz of bandwidth. Moreover, because
recording polarisation information required the EBPP to be set up in a special
mode prior to commencing observations, it is much less flexible than
\backend, which always provides full Stokes parameters for timing-mode
observations.

In addition to investigating individual observations, we also examined the
timing data of several pulsars to compare the timing stability achievable with
\backend vs. the EBPP. Depending on the pulsar, we found the weighted
root-mean-square (RMS) of the \backend timing residuals is a factor of
$\sim$1.3--3 times better than that of the EBPP over the same time interval. In our
analysis, we whitened the timing residuals with three frequency derivatives and
two DM derivatives to not be biased by the effects of pulsar spin noise and DM
variations.\footnote{The timing residuals from \backend and the EBPP
    closely trace the residuals from other EPTA telescopes, so we are confident
    that the systematic trends we are removing are not instrumental.} We also
    removed the four clock offsets affecting \backend data mentioned in
    \sect~\ref{sec:stability}. 
The smallest improvement factor we found (1.3$\times$) was for PSR~J1744$-$1134.
This is because the pulsar's particularly low DM of 3.1\,\dmunit made it
possible for the EBPP to coherently dedisperse $\sim$95\,MHz of usable
bandwidth. 

Despite \backend providing better detections than the EBPP, we still observe
with both backends in parallel whenever possible, to extend the latter's nearly
20-year long data set.

%\begin{figure}
%    \includegraphics[width=\columnwidth]{figures/ebpp_comparison/ebpp_toa_comparison.ps}
%    \caption{Comparison of TOA uncertainties, $\sigma_\mathrm{TOA}$, for
%        simultaneous observations with the \backend and EBPP backends. Equal
%        $\sigma_\mathrm{TOA}$ is shown with the dashed line. The dotted lines
%        represent 2$\times$, 5$\times$, and 10$\times$ improvements. The two
%        observations of PSR~J1713+0747 where the TOA precision is better in the
%        EBPP observations are due to there being scintillation maxima that fall
%        within the EBPP observing window. Note that we did not weight the
%        frequency channels by \snr when integrating the band.
%             \label{fig:toacompare}}
%\end{figure}

\subsection{High-Frequency Observations}
\label{sec:highfreq}
\begin{figure*}
    \includegraphics[width=\textwidth]{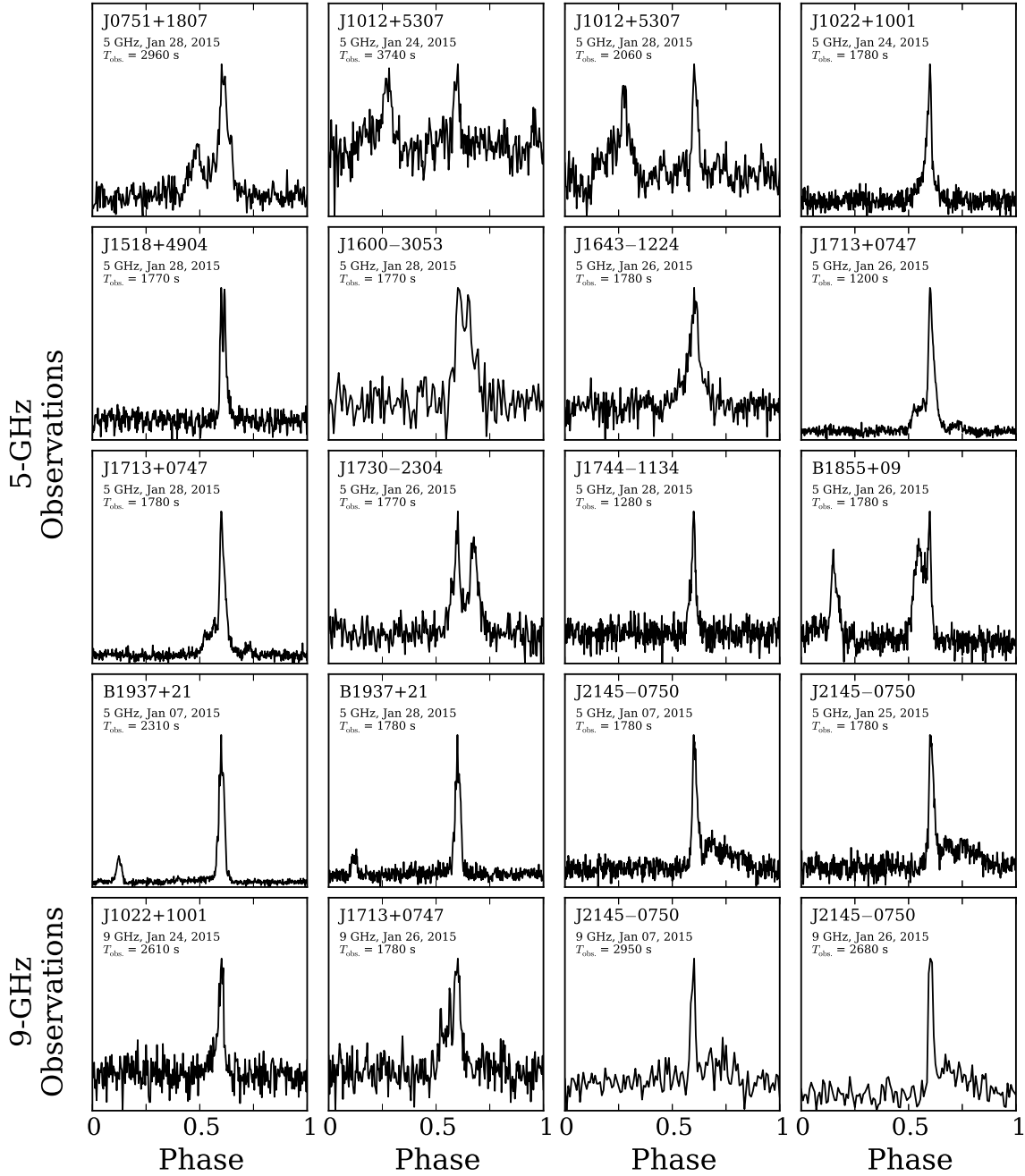}
    \caption[Pulse profiles from \backend observations at 5 and 9\,GHz]
           {Integrated pulse profiles from \backend observations at 5 and 9-GHz.
            In all cases, a bandwidth of 500\,MHz was used. These detections
            were cleaned of RFI and polarisation calibrated.  
            %Summed pulse
            %profiles from pulsars observed more than once are shown in
            %Fig.~\ref{fig:highfreq grandavg}. 
            See Table~\ref{tab:highfreq} 
            for observation details.
             \label{fig:highfreq}}
\end{figure*}

Our 5 and 9-GHz observations of 12 EPTA pulsars all resulted in detections. The
integrated pulse profiles from individual high-frequency observations are shown
in Fig.~\ref{fig:highfreq}.  PSRs~J1012+5307, J1713+0747, B1937+21 were
observed twice at 5\,GHz, and PSR~J2145$-$0750 was observed twice each 5 and
9\,GHz. 
%For these pulsars, we added observations on multiple days to create
%higher \snr pulse profiles. These are shown in Fig.~\ref{fig:highfreq
%grandavg}.  
Details of the 5 and 9-GHz observations presented in
Table~\ref{tab:highfreq}.

To flux calibrate our observations, we observed 3C~48 at both 5 and 9\,GHz
on 2015 Jan. 7 and again on 2015 Jan. 24.  These observations were used to
derive calibrated flux densities of our observations at these frequencies. We
observed the pulsars to have flux densities ranging from 0.2 to 1.5\,mJy at
5\,GHz, and from 0.2 to 0.3\,mJy at 9\,GHz (see Table~\ref{tab:highfreq}). We
found some variation in the measured flux densities of the pulsars observed
multiple times. However, these variations are consistent with amplitude
modulations expected from weak scintillation at these frequencies
\citep[e.g.][]{lk04}. 

For each of our high-frequency observations, we computed TOA uncertainties
using an analytic template, which was generated by fitting von~Mises-shaped
pulse components to the summed profile. TOA uncertainties we determined range
from 0.1 to 7.5\,$\mu$s at 5\,GHz and from 5 to 30\,$\mu$s at 9\,GHz (see
Table~\ref{tab:highfreq}).

Our 5 and 9-GHz detections show that it is feasible to monitor some MSPs at
these observing frequencies, which are higher than those typically employed for
long-term monitoring projects \citep[350\,--\,3100\,MHz; e.g. \inprep{Desvignes
et al., submitted;}][]{mhb+13,abb+15,srl+15}. It is important to note that observing
campaigns will only benefit from high-frequency detections of pulsars that are
sufficiently bright to be able to take advantage of the
reduced ISM effects. See \sect~\ref{sec:highfreqpta} for a more detailed
discussion.

%\begin{figure}
%    \includegraphics[width=\columnwidth]{figures/highfreq/highfreq_grand_averages.ps}
%    \caption[]
%        {Pulse profiles from summing multiple observations with \backend at 5
%            and 9\,GHz.  The pulse profiles from individual observations are
%            shown in Fig.~\ref{fig:highfreq}. See Table~\ref{tab:highfreq} for
%            observation details.
%             \label{fig:highfreq grandavg}}
%\end{figure}

\section{Discussion}
\label{sec:discussion}

The new \backend data set already contains roughly monthly observations of 47
MSPs, black widow pulsars and relativistic binaries at 1.4 and 2.6\,GHz
spanning at least two years, including the 33 pulsars summarised in
Tables~\ref{tab:obssumm 21cm} and~\ref{tab:obssumm 11cm}. This data set is the
successor of the venerable EBPP data set, and includes stronger detections and
more precise TOAs thanks to the larger bandwidth and more robust design of
\backend.

\subsection{Improved Sensitivity to a GW Background}
\label{sec:gwbsensitivity}

One of the primary goals of our monthly observations with \backend is to 
contribute to the EPTA and IPTA objective of detecting the GWB. To this end, we
will be combining our observations with the EPTA and IPTA data sets. 

We have estimated the improvement to GWB sensitivity made possible by switching
from the EBPP to \backend for two separate scenarios: first, assuming all
pulsars exhibit only pure white noise (e.g. radiometer noise or from pulse
jitter), and second, assuming the pulsars also suffer from red noise following a
power-law spectrum (e.g. from intrinsic spin noise or uncorrected DM
variations).  In both cases, we considered a 7-pulsar\footnote{We used the
positions of PSRs~J0218+4232, J0613$-$0200, J1022+1001, J1600$-$3053,
J1713+0747, B1855+09, and J2145$-$0750.  Because we are computing improvement
factors, we find there is no significant difference in the results as the
number of pulsars is increased.} hybrid data set that combines the higher
timing precision of the \backend TOAs with the longevity of the existing EBPP
data set.  We then compared our results for this hybrid data set to a
hypothetical extension of the current EBPP data set without having switched to
\backend. 

\begin{figure}
    \includegraphics[width=\columnwidth]{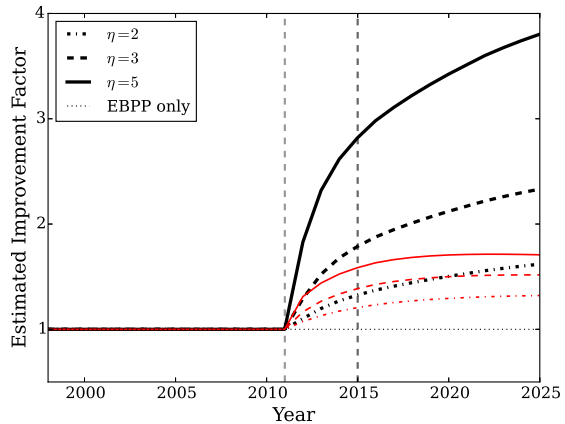}
    \caption{
        Estimated improvement in sensitivity to the GWB as a function of date
        provided by including \backend data compared to a hypothetical extended EBPP-only
        data set.  We have assumed white noise with $\rms_\mathrm{E}=1$\,$\mu$s
        for EBPP-era data (1997 to 2011), and $\rms_\mathrm{\be} =
        \rms_\mathrm{E}/\eta$ for \backend data.  The baseline of our
        comparison assumed EBPP data only. By using the Cram\'er-Rao Bound we
        calculated the GWB amplitude at which the data would show a 1\,$\sigma$
        inconsistency with the no-GWB null hypothesis.  The three black curves
        correspond to pure white noise and improvement factors of
        $\eta=2$,~3,~and~5.  The red curves include an additional source of red
        noise with an amplitude corresponding to $\rms_\mathrm{red} = 100$\,ns
        and a conservatively flat spectral index of $\alpha_\mathrm{red} =
        -1.5$ for all pulsars.  See text for additional discussion.
    \label{fig:gwbimprove}}
\end{figure}

In our first set of estimates, we assumed pulsars with pure white noise timing
residuals. We assumed the timing residuals RMS in EBPP data is $\rms_\mathrm{E}
= 1$\,$\mu$s, and the RMS of \backend residuals is improved by a factor $\eta$
(i.e. $\rms_\mathrm{\be} = \rms_\mathrm{E}/\eta$).  We performed separate
estimates for
$\eta=2$,~3,~and~5. For simplicity, we also assumed that the phase offset between the EBPP-era
data and the \backend-era data is perfectly determined.  We then used the
Cram\'er-Rao Bound \citep[e.g.][]{fis63} to compute the minimum GWB amplitude
required to reject the null hypothesis, which was that there is no GWB (i.e. a
zero-amplitude GWB) at the 1\,$\sigma$ level. A more complete description of the use of the
Cram\'er-Rao Bound in the context of estimating PTA sensitivity to the GWB can
be found in \citet{cll+15}. In making our estimates, we assumed
that the GWB signal has a power-law strain spectrum with an index of $-$2/3,
appropriate for an isotropic stochastic background of super-massive black hole
binaries.  Improvement factors were determined by comparing the GWB amplitude
derived for the hybrid data set with the analogous value computed for the pure
EBPP-style data set.  Our estimated improvement factors for $\eta=2$,~3,~and~5
as a function of date are shown in Fig.~\ref{fig:gwbimprove}.

Our second set of estimates are determined following the same procedure
described above, but assuming an additional red noise contribution to the timing
residuals.  We used a red noise spectrum with an amplitude corresponding to
$\rms_\mathrm{red} = 100$\,ns and a spectral index of $\alpha_\mathrm{red} =
-1.5$ for all pulsars. This optimistically flat value of $\alpha$ is within the
measured range for MSPs, $-7 \lapp \alpha \lapp-1$
\citep[e.g.][]{abb+15,cll+15}.\footnote{MSPs with very steep spectral indices (e.g. B1937+21) are not typically used in searches for the GWB.} Even when assuming this nearly best-case red
noise spectrum, we find the overall improvement in sensitivity to
the GWB is considerably reduced.  This is because only the power of the white
noise is reduced by switching to \backend. It is, therefore, the red noise
restricts the sensitivity to the GWB.  We also find that the improvement factor
saturates earlier because the low frequencies probed as the data set is
extended are dominated by red noise.  Thus, in the red-noise case, these lowest
frequencies contribute little sensitivity to the GWB. The improvement factors
for GWB sensitivity found for these red noise cases are indicated with the red
curves in Fig.~\ref{fig:gwbimprove}. 

%Our results are valid only in the ensemble average. Any individual realisation
%of such an experiment would show deviations due to correlations between the
%noise of the pulsars.

The difference between the black and red curves in Fig.~\ref{fig:gwbimprove} is
caused by the presence of red noise, which can arise from a variety of sources
\citep[see e.g.][]{cs10}. While it may not be possible to fully remove the
deleterious effect of red noise from pulsar timing data, some of these noise
processes (e.g. from the ISM -- see \sect~\ref{sec:highfreqpta}) can be
mitigated, further improving the prospects for the detection of the GWB.

As suggested by \citet{sejr13}, another way to counter the loss of sensitivity
to the GWB due to pulsars' red timing noise is to include other, possibly newly
discovered MSPs in PTAs. This exemplifies the importance of on-going high time
and frequency radio pulsar surveys such as the Pulsar Arecibo L-Band Feed Array
(PALFA) survey \citep{lbh+15}, the High-Time Resolution Universe (HTRU) surveys
\citep{kjv+10,bck+13}, and the Greenbank North Celestial Cap (GBNCC) survey
\citep{slr+14}.

\subsection{PTA Monitoring of MSPs at High Frequencies}
\label{sec:highfreqpta}

ISM variations, primarily DM variations, can introduce a significant amount of red noise into the timing
residuals of some MSPs \citep[e.g. \inprep{Lentati et al., in prep.;}][]{cs10,
kcs+13,lbj+14}. This can be a major hindrance to reliably detecting
long-timescale signals in the data (e.g. the nHz GWB spectrum being searched
for with PTAs).  Thus, mitigating ISM variations is of great importance. In
general, this can be accomplished in two ways: 1) by avoiding ISM variations,
either by discarding data sets contaminated by ISM variations or by observing
at frequencies high enough that the amplitude of ISM-induced noise is
sufficiently small \citep[e.g. as was done by][]{srl+15}, and 2) by removing
the ISM effects, either by leveraging multi-frequency and wide-band
observations to measure DM variations \citep[e.g.][]{kcs+13,dfg+13,abb+15}.

The effect of the ISM diminishes with increasing observing frequency:
DM delays scale as $\tau_d \propto f^{-2}$ \citep[e.g.][]{lk04} and pulse
broadening caused by interstellar scattering scales as $\tau_s \propto
f^{-3.86\pm0.16}$ \citep{bcc+04}. Therefore, pulsar timing data from
high-frequency observations will contain less significant red ISM noise.  Unfortunately,
the radio spectra of pulsars, which are generally described by a simple power
law, $S \propto f^{\alpha}$, are rather steep, with spectral indices of $-1
\lapp \alpha \lapp -2$ \citep{mkkw00,blv13}, making it difficult to completely
avoid ISM variations while maintaining the \snr required for high-precision
timing.  Thus, in practice, ISM effects cannot be completely ignored by
observing at arbitrarily high frequencies. Some effort to remove these effects
is necessary.

When removing DM variations the key resulting quantity is the infinite-frequency TOA,
$T_\infty$ (i.e. the DM-corrected TOA). Estimates of $T_\infty$ can be made by
combining multi-frequency observations or by splitting a single wide-band
observation into multiple sub-bands \citep[see e.g.][]{lbj+14}. The uncertainty
on $T_\infty$ is $\sigma_\infty = \sqrt{\left < \delta T_\infty^2 \right >}$,
and in the two-band case, is given by \citep[Eq.~12 of][]{lbj+14}

\begin{equation}
    \left < \delta T_\infty^2 \right > = \frac{f_1^4 \sigma_1^2 + f_2^4 \sigma_2^2}
                                              {\left ( f_1^2 - f_2^2 \right )^2 }, 
    \label{eq:dmcorrect}
\end{equation}

\noindent where the $f_i$ and $\sigma_i$ terms are the centre frequencies
and TOA uncertainties of the two bands, respectively.

\begin{figure}
    \includegraphics[width=\columnwidth]{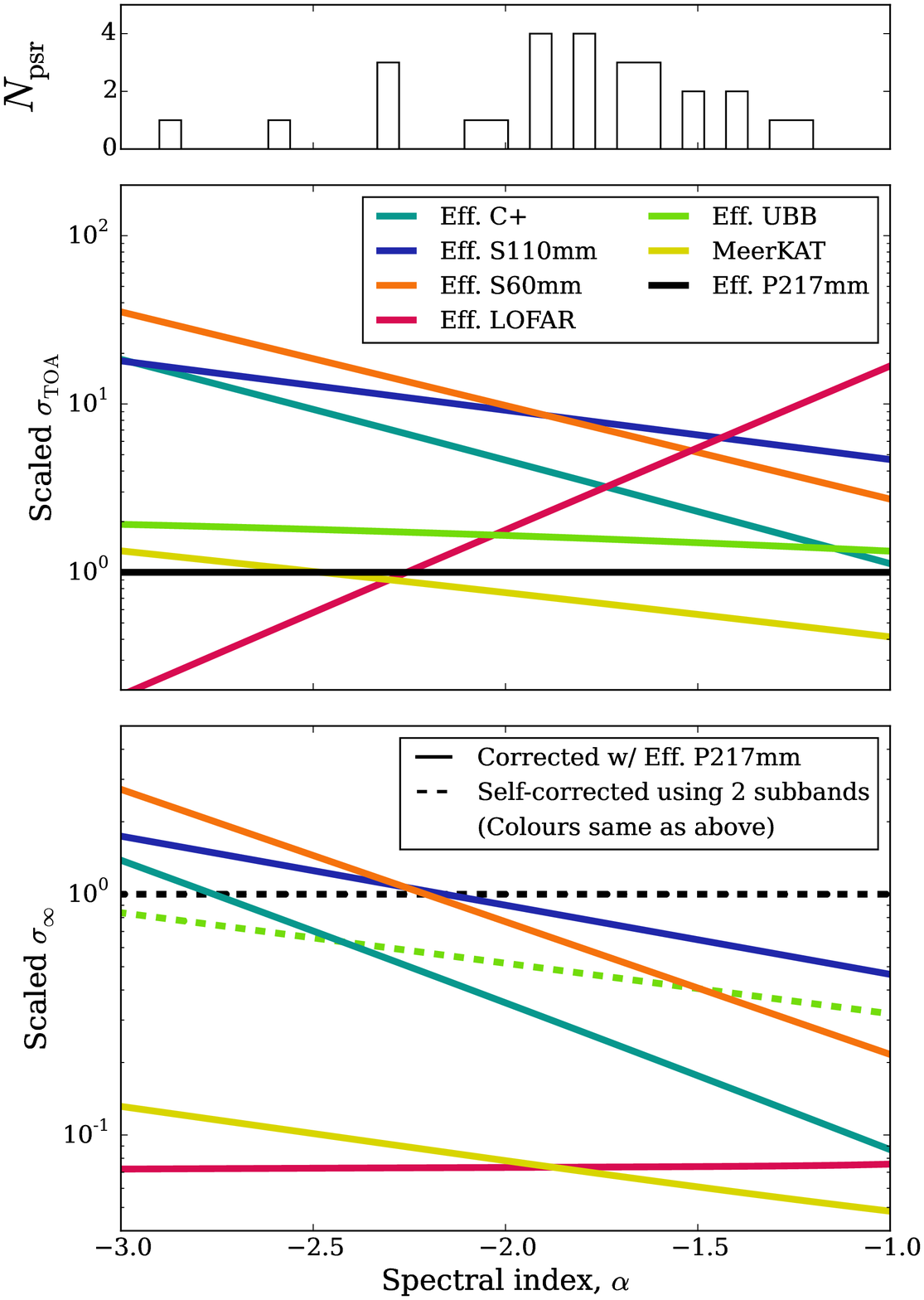}
    \caption{\textit{Top} --
             Distribution of spectral indices from the ATNF Pulsar Catalogue
             for the pulsars in Table~\ref{tab:obssumm 21cm}.
             \protect\\
             \textit{Middle} -- 
             TOA uncertainties for various existing and future observing
             systems scaled to what is achievable with \backend using the
             P217mm receiver as a function of pulsar radio spectral index.
             Lower values indicate more precise (i.e. better) TOAs. Note that
             the estimated TOA precision of the UBB receiver is worse than the
             P217mm receiver because of the former's high SEFD, which will
             be reduced when a high-pass filter is installed.
             \protect\\
             \textit{Bottom} --  
             The uncertainty of the infinite frequency TOA (i.e. the
             uncertainty of the DM-corrected TOA), $\sigma_\infty$, obtained
             from combining Effelsberg 1.4-GHz (P217mm) detections with
             data from another observing system relative to the uncertainty of
             the self-corrected P217mm observation. The $\sigma_\infty$
             values for the self-corrected observations are estimated assuming
             the band is divided evenly into two parts. Note, that doubling the
             integration time only improves $\sigma_\infty$ by $\sqrt{2}$.
             \label{fig:highfreq_comparison}
            }
\end{figure}

To measure and remove DM variations, timing data at 1~to~3\,GHz are typically
complemented by low-frequency observations (e.g. \lapp\,350\,MHz). To
illustrate the precision on $T_\infty$ attainable, we have estimated the
relative improvement in $\sigma_\infty$ from combining observations with the
LOFAR international station at Effelsberg,\footnote{The international LOFAR
    station at Effelsberg is also known as ``DE601''.} with \backend data at
    1.4\,GHz to compute. Note that we have neglected the
differences in propagation paths through Galaxy of the lower and high-frequency
radio emission due to scattering \citep[see][for a discussion of this
effect]{css15}. We estimated the ratio of TOA uncertainties derived from
observations of the same duration with different observing systems, ``A'' and
``B'', using

\begin{eqnarray}
%\mathcal{R}_\mathrm{A/B} &=& 
    \frac{\sigma_{t_A}}{\sigma_{t_B}} &=& \frac{(S/N)_B}{(S/N)_A} \nonumber \\
                         &=& \frac{S_\mathrm{sys,A}}{S_\mathrm{sys,B}} 
                             \sqrt{\frac{\Delta f_A}{\Delta f_B}} 
                             \left ( \frac{{f_{\mathrm{hi}, B}}^{(\alpha+1)} - {f_{\mathrm{lo}, B}}^{(\alpha+1)}}
                                          {{f_{\mathrm{hi}, A}}^{(\alpha+1)} - {f_{\mathrm{lo}, A}}^{(\alpha+1)}} \right ),
    \label{eq:toaunc ratio}
\end{eqnarray}

\noindent where $S_\mathrm{sys}$ is the system-equivalent flux density, $\Delta
f$ is the recorded bandwidth, $f_\mathrm{lo}$ and $f_\mathrm{hi}$ are the low
and high frequency edges of the recorded band, respectively, and $\alpha$ is
the spectral index. In deriving Eq.~\ref{eq:toaunc ratio}, we have ignored the
effect of profile evolution across the band, which is minimal for MSPs
\citep{kll+99}, as well as pulse broadening, which can be significant at
150\,MHz. We have also assumed that $S_\mathrm{sys}$ is constant across the
individual bands.  Our estimates are plotted  in
Fig.~\ref{fig:highfreq_comparison} for $-3 \leq \alpha \leq -1$.

For simplicity, in Fig.~\ref{fig:highfreq_comparison} we have assumed the
entire recorded bandwidth is summed to form a single TOA. This is a reasonable
assumption given recent work on wide-band template matching
\citep{pdr14,ldc+14}. In fact, by using these new wide-band TOA determination
algorithms it is possible to simultaneously account for profile evolution, DM
variations, scattering, and scintillation while summarising wide-band
observations into a single TOA. 

Clearly, complementing \backend TOAs with observations using the Effelsberg
LOFAR station, provides precise DM-corrected TOAs.  However, there are some
complications with using low-frequency data to remove DM variations.  First,
the background of Galactic synchrotron emission is strong and line-of-sight
dependent \citep{hssw82}. Thus, MSPs in unfortunate directions may be too weak
to use low-frequency observations to make DM measurements.  Another possible
complication, not included in our estimates, is spectral turnover, which occurs
in a significant fraction of MSPs \citep{kvl+15}. This also conspires to weaken
detections at low frequencies.  Finally, there is concern that the DMs measured
at low frequencies are different than those measured at higher
frequencies due to differences in the ISM probed as a result of interstellar
scattering \citep{css15}.

Our 5 and 9-GHz detections of several MSPs indicate that complementing
timing campaigns at 1 to 3\,GHz with observations at higher frequencies might
be a viable alternative for mitigating ISM effects.
Fig.~\ref{fig:highfreq_comparison} includes estimates of $\sigma_\infty$
attainable by complementing 1.4\,GHz \backend data with observations from
current and planned receiver systems. Using the current 5-GHz
set-up with \backend should provide better TOA precision than the 2.6-GHz
set-up described in \sect~\ref{sec:observations} for $\alpha \gapp -1.9$, % measured from plot
as well as better $\sigma_\infty$ for $\alpha \gapp -2.25$, thanks to the larger
frequency separation between the bands. 

We also examined new wide-bandwidth receivers coming online, such as the
0.6-to-2.7-GHz UBB receiver at
Effelsberg,\footnote{Currently, the UBB receiver at Effelsberg can observe
frequencies as low as 0.6\,GHz. However, since the RFI is particularly
strong below 1\,GHz, a high-pass filter will be installed around 1.1\,GHz to
improve the overall performance of the receiver.} as well as the new 4-to-9-GHz ``C+'' 
receiver currently being commissioned at
Effelsberg, and the 1.8-to-3.5-GHz receivers being designed at the MPIfR for
MeerKAT.  The assumed receiver parameters are
shown Table~\ref{tab:otherrcvr}. 

Effelsberg's C+ receiver and the MeerKAT 1.8-to-3.5-GHz observing systems
present significant improvements in the bandwidths currently available for
pulsar timing at these frequencies, and thus provide compelling cases for using
high-frequency observations to mitigate ISM-related noise. Furthermore, other
sensitive high-frequency observing systems (e.g. the 3.85--6\,GHz ``C-band''
receiver at the Arecibo Observatory, the 2--4\,GHz ``S-band'' system at the
Jansky Very Large Array (JVLA), and the 4--8\,GHz ``C-band'' system at the
JVLA) should be considered for regular PTA monitoring of MSPs. Furthermore, the
potential high-frequency capabilities of the SKA could also be important for
PTA-studies.

Note that, at $\sim$5\,GHz we expect slow, roughly hourly, intensity variations
of only $\sim$10~to~30\,\%. Also, the attenuation of the transmission of radio waves
through the atmosphere due to water vapour only becomes a concern at
$f\gapp$10\,GHz.

%Finally, pulsar profiles are generally narrower at higher observing frequencies
%\citep{cor78}. In principle, observing at higher frequencies results in sharper
%pulse profiles, which enables more precise TOA determination. However, this
%narrowing effect is less pronounced for MSPs \citep{kll+99}.  

%Fortunately, this effect (attenuation at $f > 10$\,GHz due to water
%vapour/weather) can be predicted, and plans can be altered to observe at lower
%frequencies on days when the weather is not suitable for such high frequency
%observing, especially at a telescope like Effelsberg, which can switch between
%its 2-GHz, 5-GHz, and 9-GHz secondary focus receivers in only $\sim 30$\,s.

When formulating observing strategies for high-precision,
multi-frequency timing campaigns, deciding what observing systems to use
should depend on the spectral index of the pulsar, the magnitude of the noise
introduced by ISM effects, and the shape of the pulse profile, including the
degree of profile broadening. Furthermore, deciding how often multi-frequency
observations are necessary for a particular pulsar depends on the timescale of
the DM variations affecting the pulsar \citep[see e.g.][]{kcs+13}.  At
Effelsberg, because the secondary-focus receivers are always available, and can
be switched-to within minutes, issues arising from the non-simultaneity of
multi-frequency observations described by \citet{lccd15} are not a concern, as
they may be at other telescopes. 

\section{Conclusions}
\label{sec:conclusion}
We have described how the \backend backend is being used at the Effelsberg
radio telescope for monthly EPTA observations. The coherently dedispersed data
from \backend have a larger bandwidth than what is possible with its
predecessor, the EBPP. As a result, the now four-year-old \backend data set has
stronger detections, more precise TOAs, and will improve the sensitivity to the
GWB compared to the EBPP data.

We have also shown how bright, highly stable MSPs being monitored by the EPTA
can be detected at 5 and 9\,GHz. Given the reduced ISM effects at these
frequencies compared to 1.4\,GHz and the ability to more precisely correct DM
variations, there could be advantages to complementing existing data sets
(typically $1 \lapp f \lapp 3$\,GHz) with observations at these higher
frequencies.  This is especially true considering the new wide-bandwidth,
high-frequency receivers currently being commissioned, and those expected to
come online in the next few years.
\\

\noindent We would like to thank Cees Bassa and Jason Hessels for helping to
conceive and develop \toaster. We would also like to thank Paulo Freire and
Gregory Desvignes for useful discussions, as well as Stefan Os{\l}owski, Golam
Shaifullah, Antoine Lassus, and John Antoniadis for helping to observe. PL and
EG acknowledge support from IMPRS Bonn/Cologne. RNC acknowledges the support of
IMPRS Bonn/Cologne and the Bonn-Cologne Graduate School.  KJL gratefully
acknowledges support from National Basic Research Program of China, 973
Program, 2015CB857101 and NSFC 11373011.  The \backend backend was constructed
as part of the LEAP project, which was funded by the ERC Advanced Grant “LEAP”,
Grant Agreement Number 227947 (PI M. Kramer).

\bibliographystyle{apj}
\bibliography{asterix}

\begin{thebibliography}{}
\expandafter\ifx\csname natexlab\endcsname\relax\def\natexlab#1{#1}\fi

\bibitem[{{Antoniadis} {et~al.}(2013){Antoniadis}, {Freire}, {Wex}, {Tauris},
  {Lynch}, {van Kerkwijk}, {Kramer}, {Bassa}, {Dhillon}, {Driebe}, {Hessels},
  {Kaspi}, {Kondratiev}, {Langer}, {Marsh}, {McLaughlin}, {Pennucci}, {Ransom},
  {Stairs}, {van Leeuwen}, {Verbiest}, \& {Whelan}}]{afw+13}
{Antoniadis}, J., {Freire}, P.~C.~C., {Wex}, N., {et~al.} 2013, Science, 340,
  448

\bibitem[{{Arzoumanian} {et~al.}(2015){Arzoumanian}, {Brazier},
  {Burke-Spolaor}, {Chamberlin}, {Chatterjee}, {Christy}, {Cordes}, {Cornish},
  {Crowter}, {Demorest}, {Dolch}, {Ellis}, {Ferdman}, {Fonseca},
  {Garver-Daniels}, {Gonzalez}, {Jenet}, {Jones}, {Jones}, {Kaspi}, {Koop},
  {Lazio}, {Lam}, {Levin}, {Lommen}, {Lorimer}, {Luo}, {Lynch}, {Madison},
  {McLaughlin}, {McWilliams}, {Nice}, {Palliyaguru}, {Pennucci}, {Ransom},
  {Siemens}, {Stairs}, {Stinebring}, {Stovall}, {Swiggum}, {Vallisneri}, {van
  Haasteren}, {Wang}, \& {Zhu}}]{abb+15}
{Arzoumanian}, Z., {Brazier}, A., {Burke-Spolaor}, S., {et~al.} 2015, ArXiv
  e-prints, arXiv:1505.07540

\bibitem[{{Backer} {et~al.}(1997){Backer}, {Dexter}, {Zepka}, {Ng},
  {Werthimer}, {Ray}, \& {Foster}}]{bdz+97}
{Backer}, D.~C., {Dexter}, M.~R., {Zepka}, A., {et~al.} 1997, \pasp, 109, 61

\bibitem[{{Barr} {et~al.}(2013){Barr}, {Champion}, {Kramer}, {Eatough},
  {Freire}, {Karuppusamy}, {Lee}, {Verbiest}, {Bassa}, {Lyne}, {Stappers},
  {Lorimer}, \& {Klein}}]{bck+13}
{Barr}, E.~D., {Champion}, D.~J., {Kramer}, M., {et~al.} 2013, \mnras, 435,
  2234

\bibitem[{{Bassa} {et~al.}(2015){Bassa}, {Janssen}, {Karuppusamy}, {Kramer},
  {Lee}, {Liu}, {McKee}, {Perrodin}, {Purver}, {Sanidas}, {Smits}, \&
  {Stappers}}]{bjk+15}
{Bassa}, C.~G., {Janssen}, G.~H., {Karuppusamy}, R., {et~al.} 2015, ArXiv
  e-prints, arXiv:1511.06597

\bibitem[{{Bates} {et~al.}(2013){Bates}, {Lorimer}, \& {Verbiest}}]{blv13}
{Bates}, S.~D., {Lorimer}, D.~R., \& {Verbiest}, J.~P.~W. 2013, \mnras, 431,
  1352

\bibitem[{{Berkhuijsen} \& {M{\"u}ller}(2008)}]{bm08}
{Berkhuijsen}, E.~M., \& {M{\"u}ller}, P. 2008, \aap, 490, 179

\bibitem[{{Bhat} {et~al.}(2004){Bhat}, {Cordes}, {Camilo}, {Nice}, \&
  {Lorimer}}]{bcc+04}
{Bhat}, N.~D.~R., {Cordes}, J.~M., {Camilo}, F., {Nice}, D.~J., \& {Lorimer},
  D.~R. 2004, \apj, 605, 759

\bibitem[{{Bhat} {et~al.}(1998){Bhat}, {Gupta}, \& {Rao}}]{bgr98}
{Bhat}, N.~D.~R., {Gupta}, Y., \& {Rao}, A.~P. 1998, \apj, 500, 262

\bibitem[{{Britton}(2000)}]{bri00}
{Britton}, M.~C. 2000, \apj, 532, 1240

\bibitem[{{Caballero} {et~al.}(2015){Caballero}, {Lee}, {Lentati}, {Desvignes},
  {Champion}, {Verbiest}, {Janssen}, {Stappers}, {Kramer}, {Lazarus},
  {Possenti}, {Tiburzi}, {Perrodin}, {Os{\l}owski}, {Babak}, {Bassa}, {Brem},
  {Burgay}, {Cognard}, {Gair}, {Graikou}, {Guillemot}, {Hessels},
  {Karuppusamy}, {Lassus}, {Liu}, {McKee}, {Mingarelli}, {Petiteau}, {Purver},
  {Rosado}, {Sanidas}, {Sesana}, {Shaifullah}, {Smits}, {Taylor}, {Theureau},
  {van Haasteren}, \& {Vecchio}}]{cll+15}
{Caballero}, R.~N., {Lee}, K.~J., {Lentati}, L., {et~al.} 2015, ArXiv e-prints,
  arXiv:1510.09194

\bibitem[{{Cognard} {et~al.}(2011){Cognard}, {Guillemot}, {Johnson}, {Smith},
  {Venter}, {Harding}, {Wolff}, {Cheung}, {Donato}, {Abdo}, {Ballet}, {Camilo},
  {Desvignes}, {Dumora}, {Ferrara}, {Freire}, {Grove}, {Johnston}, {Keith},
  {Kramer}, {Lyne}, {Michelson}, {Parent}, {Ransom}, {Ray}, {Romani}, {Saz
  Parkinson}, {Stappers}, {Theureau}, {Thompson}, {Weltevrede}, \&
  {Wood}}]{cgj+11}
{Cognard}, I., {Guillemot}, L., {Johnson}, T.~J., {et~al.} 2011, \apj, 732, 47

\bibitem[{{Cordes} \& {Shannon}(2010)}]{cs10}
{Cordes}, J.~M., \& {Shannon}, R.~M. 2010, ArXiv e-prints, arXiv:1010.3785

\bibitem[{{Cordes} {et~al.}(2015){Cordes}, {Shannon}, \& {Stinebring}}]{css15}
{Cordes}, J.~M., {Shannon}, R.~M., \& {Stinebring}, D.~R. 2015, ArXiv e-prints,
  arXiv:1503.08491

\bibitem[{{Demorest} {et~al.}(2010){Demorest}, {Pennucci}, {Ransom}, {Roberts},
  \& {Hessels}}]{dpr+10}
{Demorest}, P.~B., {Pennucci}, T., {Ransom}, S.~M., {Roberts}, M.~S.~E., \&
  {Hessels}, J.~W.~T. 2010, \nat, 467, 1081

\bibitem[{{Demorest} {et~al.}(2013){Demorest}, {Ferdman}, {Gonzalez}, {Nice},
  {Ransom}, {Stairs}, {Arzoumanian}, {Brazier}, {Burke-Spolaor}, {Chamberlin},
  {Cordes}, {Ellis}, {Finn}, {Freire}, {Giampanis}, {Jenet}, {Kaspi}, {Lazio},
  {Lommen}, {McLaughlin}, {Palliyaguru}, {Perrodin}, {Shannon}, {Siemens},
  {Stinebring}, {Swiggum}, \& {Zhu}}]{dfg+13}
{Demorest}, P.~B., {Ferdman}, R.~D., {Gonzalez}, M.~E., {et~al.} 2013, \apj,
  762, 94

\bibitem[{{Dolch} {et~al.}(2014){Dolch}, {Lam}, {Cordes}, {Chatterjee},
  {Bassa}, {Bhattacharyya}, {Champion}, {Cognard}, {Crowter}, {Demorest},
  {Hessels}, {Janssen}, {Jenet}, {Jones}, {Jordan}, {Karuppusamy}, {Keith},
  {Kondratiev}, {Kramer}, {Lazarus}, {Lazio}, {Lee}, {McLaughlin}, {Roy},
  {Shannon}, {Stairs}, {Stovall}, {Verbiest}, {Madison}, {Palliyaguru},
  {Perrodin}, {Ransom}, {Stappers}, {Zhu}, {Dai}, {Desvignes}, {Guillemot},
  {Liu}, {Lyne}, {Perera}, {Petroff}, {Rankin}, \& {Smits}}]{dlc+14}
{Dolch}, T., {Lam}, M.~T., {Cordes}, J., {et~al.} 2014, \apj, 794, 21

\bibitem[{{Eatough} {et~al.}(2013){Eatough}, {Falcke}, {Karuppusamy}, {Lee},
  {Champion}, {Keane}, {Desvignes}, {Schnitzeler}, {Spitler}, {Kramer},
  {Klein}, {Bassa}, {Bower}, {Brunthaler}, {Cognard}, {Deller}, {Demorest},
  {Freire}, {Kraus}, {Lyne}, {Noutsos}, {Stappers}, \& {Wex}}]{efk+13}
{Eatough}, R.~P., {Falcke}, H., {Karuppusamy}, R., {et~al.} 2013, \nat, 501,
  391

\bibitem[{Fisz(1963)}]{fis63}
Fisz, M. 1963, Probability theory and mathematical statistics (Polish
  Scientifique)

\bibitem[{{Ford} {et~al.}(2010){Ford}, {Demorest}, \& {Ransom}}]{fdr10}
{Ford}, J.~M., {Demorest}, P., \& {Ransom}, S. 2010, in Society of
  Photo-Optical Instrumentation Engineers (SPIE) Conference Series, Vol. 7740,
  Society of Photo-Optical Instrumentation Engineers (SPIE) Conference Series,
  0

\bibitem[{{Freire} {et~al.}(2011){Freire}, {Bassa}, {Wex}, {Stairs},
  {Champion}, {Ransom}, {Lazarus}, {Kaspi}, {Hessels}, {Kramer}, {Cordes},
  {Verbiest}, {Podsiadlowski}, {Nice}, {Deneva}, {Lorimer}, {Stappers},
  {McLaughlin}, \& {Camilo}}]{fbw+11}
{Freire}, P.~C.~C., {Bassa}, C.~G., {Wex}, N., {et~al.} 2011, \mnras, 412, 2763

\bibitem[{{Haslam} {et~al.}(1982){Haslam}, {Salter}, {Stoffel}, \&
  {Wilson}}]{hssw82}
{Haslam}, C.~G.~T., {Salter}, C.~J., {Stoffel}, H., \& {Wilson}, W.~E. 1982,
  \aaps, 47, 1

\bibitem[{{Hobbs} {et~al.}(2006){Hobbs}, {Edwards}, \& {Manchester}}]{hem06}
{Hobbs}, G.~B., {Edwards}, R.~T., \& {Manchester}, R.~N. 2006, \mnras, 369, 655

\bibitem[{{Hotan} {et~al.}(2004){Hotan}, {van Straten}, \&
  {Manchester}}]{hvm04}
{Hotan}, A.~W., {van Straten}, W., \& {Manchester}, R.~N. 2004, \pasa, 21, 302

\bibitem[{{Jacoby} {et~al.}(2007){Jacoby}, {Bailes}, {Ord}, {Knight}, \&
  {Hotan}}]{jbo+07}
{Jacoby}, B.~A., {Bailes}, M., {Ord}, S.~M., {Knight}, H.~S., \& {Hotan}, A.~W.
  2007, \apj, 656, 408

\bibitem[{{Janssen} {et~al.}(2010){Janssen}, {Stappers}, {Bassa}, {Cognard},
  {Kramer}, \& {Theureau}}]{jsb+10}
{Janssen}, G.~H., {Stappers}, B.~W., {Bassa}, C.~G., {et~al.} 2010, \aap, 514,
  A74

\bibitem[{{Janssen} {et~al.}(2008){Janssen}, {Stappers}, {Kramer}, {Nice},
  {Jessner}, {Cognard}, \& {Purver}}]{jsk+08}
{Janssen}, G.~H., {Stappers}, B.~W., {Kramer}, M., {et~al.} 2008, \aap, 490,
  753

\bibitem[{{Keith} {et~al.}(2010){Keith}, {Jameson}, {van Straten}, {Bailes},
  {Johnston}, {Kramer}, {Possenti}, {Bates}, {Bhat}, {Burgay}, {Burke-Spolaor},
  {D'Amico}, {Levin}, {McMahon}, {Milia}, \& {Stappers}}]{kjv+10}
{Keith}, M.~J., {Jameson}, A., {van Straten}, W., {et~al.} 2010, \mnras, 409,
  619

\bibitem[{{Keith} {et~al.}(2013){Keith}, {Coles}, {Shannon}, {Hobbs},
  {Manchester}, {Bailes}, {Bhat}, {Burke-Spolaor}, {Champion}, {Chaudhary},
  {Hotan}, {Khoo}, {Kocz}, {Os{\l}owski}, {Ravi}, {Reynolds}, {Sarkissian},
  {van Straten}, \& {Yardley}}]{kcs+13}
{Keith}, M.~J., {Coles}, W., {Shannon}, R.~M., {et~al.} 2013, \mnras, 429, 2161

\bibitem[{{Kondratiev} {et~al.}(2015){Kondratiev}, {Verbiest}, {Hessels},
  {Bilous}, {Stappers}, {Kramer}, {Keane}, {Noutsos}, {Os{\l}owski}, {Breton},
  {Hassall}, {Alexov}, {Cooper}, {Falcke}, {Grie{\ss}meier}, {Karastergiou},
  {Kuniyoshi}, {Pilia}, {Sobey}, {ter Veen}, {Weltevrede}, {Bell}, {Broderick},
  {Corbel}, {Eisl{\"o}ffel}, {Markoff}, {Rowlinson}, {Swinbank}, {Wijers},
  {Wijnands}, \& {Zarka}}]{kvh+15}
{Kondratiev}, V.~I., {Verbiest}, J.~P.~W., {Hessels}, J.~W.~T., {et~al.} 2015,
  ArXiv e-prints, arXiv:1508.02948

\bibitem[{{Kramer} {et~al.}(1999){Kramer}, {Lange}, {Lorimer}, {Backer},
  {Xilouris}, {Jessner}, \& {Wielebinski}}]{kll+99}
{Kramer}, M., {Lange}, C., {Lorimer}, D.~R., {et~al.} 1999, \apj, 526, 957

\bibitem[{{Kramer} {et~al.}(1998){Kramer}, {Xilouris}, {Lorimer}, {Doroshenko},
  {Jessner}, {Wielebinski}, {Wolszczan}, \& {Camilo}}]{kxl+98}
{Kramer}, M., {Xilouris}, K.~M., {Lorimer}, D.~R., {et~al.} 1998, \apj, 501,
  270

\bibitem[{{Kramer} {et~al.}(2006){Kramer}, {Stairs}, {Manchester},
  {McLaughlin}, {Lyne}, {Ferdman}, {Burgay}, {Lorimer}, {Possenti}, {D'Amico},
  {Sarkissian}, {Hobbs}, {Reynolds}, {Freire}, \& {Camilo}}]{ksm+06}
{Kramer}, M., {Stairs}, I.~H., {Manchester}, R.~N., {et~al.} 2006, Science,
  314, 97

\bibitem[{{Kuniyoshi} {et~al.}(2015){Kuniyoshi}, {Verbiest}, {Lee}, {Adebahr},
  {Kramer}, \& {Noutsos}}]{kvl+15}
{Kuniyoshi}, M., {Verbiest}, J.~P.~W., {Lee}, K.~J., {et~al.} 2015, ArXiv
  e-prints, arXiv:1507.03732

\bibitem[{{Lam} {et~al.}(2015){Lam}, {Cordes}, {Chatterjee}, \&
  {Dolch}}]{lccd15}
{Lam}, M.~T., {Cordes}, J.~M., {Chatterjee}, S., \& {Dolch}, T. 2015, \apj,
  801, 130

\bibitem[{{Lazaridis} {et~al.}(2009){Lazaridis}, {Wex}, {Jessner}, {Kramer},
  {Stappers}, {Janssen}, {Desvignes}, {Purver}, {Cognard}, {Theureau}, {Lyne},
  {Jordan}, \& {Zensus}}]{lwj+09}
{Lazaridis}, K., {Wex}, N., {Jessner}, A., {et~al.} 2009, \mnras, 400, 805

\bibitem[{{Lazaridis} {et~al.}(2011){Lazaridis}, {Verbiest}, {Tauris},
  {Stappers}, {Kramer}, {Wex}, {Jessner}, {Cognard}, {Desvignes}, {Janssen},
  {Purver}, {Theureau}, {Bassa}, \& {Smits}}]{lvt+11}
{Lazaridis}, K., {Verbiest}, J.~P.~W., {Tauris}, T.~M., {et~al.} 2011, \mnras,
  414, 3134

\bibitem[{{Lazarus} {et~al.}(2015){Lazarus}, {Brazier}, {Hessels},
  {Karako-Argaman}, {Kaspi}, {Lynch}, {Madsen}, {Patel}, {Ransom}, {Scholz},
  {Swiggum}, {Zhu}, {Allen}, {Bogdanov}, {Camilo}, {Cardoso}, {Chatterjee},
  {Cordes}, {Crawford}, {Deneva}, {Ferdman}, {Freire}, {Jenet}, {Knispel},
  {Lee}, {van Leeuwen}, {Lorimer}, {Lyne}, {McLaughlin}, {Siemens}, {Spitler},
  {Stairs}, {Stovall}, \& {Venkataraman}}]{lbh+15}
{Lazarus}, P., {Brazier}, A., {Hessels}, J.~W.~T., {et~al.} 2015, \apj, 812, 81

\bibitem[{{Lee} {et~al.}(2014){Lee}, {Bassa}, {Janssen}, {Karuppusamy},
  {Kramer}, {Liu}, {Perrodin}, {Smits}, {Stappers}, {van Haasteren}, \&
  {Lentati}}]{lbj+14}
{Lee}, K.~J., {Bassa}, C.~G., {Janssen}, G.~H., {et~al.} 2014, \mnras, 441,
  2831

\bibitem[{{Lentati} {et~al.}(2014){Lentati}, {Alexander}, {Hobson}, {Feroz},
  {van Haasteren}, {Lee}, \& {Shannon}}]{lah+14}
{Lentati}, L., {Alexander}, P., {Hobson}, M.~P., {et~al.} 2014, \mnras, 437,
  3004

\bibitem[{{Lentati} {et~al.}(2015){Lentati}, {Taylor}, {Mingarelli}, {Sesana},
  {Sanidas}, {Vecchio}, {Caballero}, {Lee}, {van Haasteren}, {Babak}, {Bassa},
  {Brem}, {Burgay}, {Champion}, {Cognard}, {Desvignes}, {Gair}, {Guillemot},
  {Hessels}, {Janssen}, {Karuppusamy}, {Kramer}, {Lassus}, {Lazarus}, {Liu},
  {Os{\l}owski}, {Perrodin}, {Petiteau}, {Possenti}, {Purver}, {Rosado},
  {Smits}, {Stappers}, {Theureau}, {Tiburzi}, \& {Verbiest}}]{ltm+15}
{Lentati}, L., {Taylor}, S.~R., {Mingarelli}, C.~M.~F., {et~al.} 2015, ArXiv
  e-prints, arXiv:1504.03692

\bibitem[{{Liu} {et~al.}(2014){Liu}, {Desvignes}, {Cognard}, {Stappers},
  {Verbiest}, {Lee}, {Champion}, {Kramer}, {Freire}, \& {Karuppusamy}}]{ldc+14}
{Liu}, K., {Desvignes}, G., {Cognard}, I., {et~al.} 2014, \mnras, 443, 3752

\bibitem[{{Lommen} {et~al.}(2000){Lommen}, {Zepka}, {Backer}, {McLaughlin},
  {Cordes}, {Arzoumanian}, \& {Xilouris}}]{lzb+00}
{Lommen}, A.~N., {Zepka}, A., {Backer}, D.~C., {et~al.} 2000, \apj, 545, 1007

\bibitem[{{Lorimer} \& {Kramer}(2004)}]{lk04}
{Lorimer}, D.~R., \& {Kramer}, M. 2004, {Handbook of Pulsar Astronomy}, ed.
  R.~{Ellis}, J.~{Huchra}, S.~{Kahn}, G.~{Rieke}, \& P.~B. {Stetson}

\bibitem[{{Lorimer} {et~al.}(2006){Lorimer}, {Faulkner}, {Lyne}, {Manchester},
  {Kramer}, {McLaughlin}, {Hobbs}, {Possenti}, {Stairs}, {Camilo}, {Burgay},
  {D'Amico}, {Corongiu}, \& {Crawford}}]{lfl+06}
{Lorimer}, D.~R., {Faulkner}, A.~J., {Lyne}, A.~G., {et~al.} 2006, \mnras, 372,
  777

\bibitem[{{Lynch} {et~al.}(2013){Lynch}, {Boyles}, {Ransom}, {Stairs},
  {Lorimer}, {McLaughlin}, {Hessels}, {Kaspi}, {Kondratiev}, {Archibald},
  {Berndsen}, {Cardoso}, {Cherry}, {Epstein}, {Karako-Argaman}, {McPhee},
  {Pennucci}, {Roberts}, {Stovall}, \& {van Leeuwen}}]{lbr+13}
{Lynch}, R.~S., {Boyles}, J., {Ransom}, S.~M., {et~al.} 2013, \apj, 763, 81

\bibitem[{{Manchester} {et~al.}(2013){Manchester}, {Hobbs}, {Bailes}, {Coles},
  {van Straten}, {Keith}, {Shannon}, {Bhat}, {Brown}, {Burke-Spolaor},
  {Champion}, {Chaudhary}, {Edwards}, {Hampson}, {Hotan}, {Jameson}, {Jenet},
  {Kesteven}, {Khoo}, {Kocz}, {Maciesiak}, {Oslowski}, {Ravi}, {Reynolds},
  {Sarkissian}, {Verbiest}, {Wen}, {Wilson}, {Yardley}, {Yan}, \&
  {You}}]{mhb+13}
{Manchester}, R.~N., {Hobbs}, G., {Bailes}, M., {et~al.} 2013, \pasa, 30, 17

\bibitem[{{Maron} {et~al.}(2000){Maron}, {Kijak}, {Kramer}, \&
  {Wielebinski}}]{mkkw00}
{Maron}, O., {Kijak}, J., {Kramer}, M., \& {Wielebinski}, R. 2000, \aaps, 147,
  195

\bibitem[{{Ng} {et~al.}(2014){Ng}, {Bailes}, {Bates}, {Bhat}, {Burgay},
  {Burke-Spolaor}, {Champion}, {Coster}, {Johnston}, {Keith}, {Kramer},
  {Levin}, {Petroff}, {Possenti}, {Stappers}, {van Straten}, {Thornton},
  {Tiburzi}, {Bassa}, {Freire}, {Guillemot}, {Lyne}, {Tauris}, {Shannon}, \&
  {Wex}}]{cbb+14}
{Ng}, C., {Bailes}, M., {Bates}, S.~D., {et~al.} 2014, \mnras, 439, 1865

\bibitem[{{Pennucci} {et~al.}(2014){Pennucci}, {Demorest}, \& {Ransom}}]{pdr14}
{Pennucci}, T.~T., {Demorest}, P.~B., \& {Ransom}, S.~M. 2014, \apj, 790, 93

\bibitem[{{Sanidas} {et~al.}(2012){Sanidas}, {Battye}, \& {Stappers}}]{sbs12}
{Sanidas}, S.~A., {Battye}, R.~A., \& {Stappers}, B.~W. 2012, \prd, 85, 122003

\bibitem[{{Sesana}(2013)}]{ses13}
{Sesana}, A. 2013, Classical and Quantum Gravity, 30, 224014

\bibitem[{{Shannon} {et~al.}(2015){Shannon}, {Ravi}, {Lentati}, {Lasky},
  {Hobbs}, {Kerr}, {Manchester}, {Coles}, {Levin}, {Bailes}, {Bhat},
  {Burke-Spolaor}, {Dai}, {Keith}, {Os{\l}owski}, {Reardon}, {van Straten},
  {Toomey}, {Wang}, {Wen}, {Wyithe}, \& {Zhu}}]{srl+15}
{Shannon}, R.~M., {Ravi}, V., {Lentati}, L.~T., {et~al.} 2015, Science, 349,
  1522

\bibitem[{{Shao} {et~al.}(2013){Shao}, {Caballero}, {Kramer}, {Wex},
  {Champion}, \& {Jessner}}]{sck+13}
{Shao}, L., {Caballero}, R.~N., {Kramer}, M., {et~al.} 2013, Classical and
  Quantum Gravity, 30, 165019

\bibitem[{{Siemens} {et~al.}(2013){Siemens}, {Ellis}, {Jenet}, \&
  {Romano}}]{sejr13}
{Siemens}, X., {Ellis}, J., {Jenet}, F., \& {Romano}, J.~D. 2013, Classical and
  Quantum Gravity, 30, 224015

\bibitem[{{Stairs} {et~al.}(2005){Stairs}, {Faulkner}, {Lyne}, {Kramer},
  {Lorimer}, {McLaughlin}, {Manchester}, {Hobbs}, {Camilo}, {Possenti},
  {Burgay}, {D'Amico}, {Freire}, \& {Gregory}}]{sfl+05}
{Stairs}, I.~H., {Faulkner}, A.~J., {Lyne}, A.~G., {et~al.} 2005, \apj, 632,
  1060

\bibitem[{{Stovall} {et~al.}(2014){Stovall}, {Lynch}, {Ransom}, {Archibald},
  {Banaszak}, {Biwer}, {Boyles}, {Dartez}, {Day}, {Ford}, {Flanigan}, {Garcia},
  {Hessels}, {Hinojosa}, {Jenet}, {Kaplan}, {Karako-Argaman}, {Kaspi},
  {Kondratiev}, {Leake}, {Lorimer}, {Lunsford}, {Martinez}, {Mata},
  {McLaughlin}, {Roberts}, {Rohr}, {Siemens}, {Stairs}, {van Leeuwen},
  {Walker}, \& {Wells}}]{slr+14}
{Stovall}, K., {Lynch}, R.~S., {Ransom}, S.~M., {et~al.} 2014, \apj, 791, 67

\bibitem[{{van Haarlem} {et~al.}(2013){van Haarlem}, {Wise}, {Gunst}, {Heald},
  {McKean}, {Hessels}, {de Bruyn}, {Nijboer}, {Swinbank}, {Fallows},
  {Brentjens}, {Nelles}, {Beck}, {Falcke}, {Fender}, {H{\"o}randel},
  {Koopmans}, {Mann}, {Miley}, {R{\"o}ttgering}, {Stappers}, {Wijers},
  {Zaroubi}, {van den Akker}, {Alexov}, {Anderson}, {Anderson}, {van Ardenne},
  {Arts}, {Asgekar}, {Avruch}, {Batejat}, {B{\"a}hren}, {Bell}, {Bell}, {van
  Bemmel}, {Bennema}, {Bentum}, {Bernardi}, {Best}, {B{\^i}rzan}, {Bonafede},
  {Boonstra}, {Braun}, {Bregman}, {Breitling}, {van de Brink}, {Broderick},
  {Broekema}, {Brouw}, {Br{\"u}ggen}, {Butcher}, {van Cappellen}, {Ciardi},
  {Coenen}, {Conway}, {Coolen}, {Corstanje}, {Damstra}, {Davies}, {Deller},
  {Dettmar}, {van Diepen}, {Dijkstra}, {Donker}, {Doorduin}, {Dromer}, {Drost},
  {van Duin}, {Eisl{\"o}ffel}, {van Enst}, {Ferrari}, {Frieswijk}, {Gankema},
  {Garrett}, {de Gasperin}, {Gerbers}, {de Geus}, {Grie{\ss}meier}, {Grit},
  {Gruppen}, {Hamaker}, {Hassall}, {Hoeft}, {Holties}, {Horneffer}, {van der
  Horst}, {van Houwelingen}, {Huijgen}, {Iacobelli}, {Intema}, {Jackson},
  {Jelic}, {de Jong}, {Juette}, {Kant}, {Karastergiou}, {Koers}, {Kollen},
  {Kondratiev}, {Kooistra}, {Koopman}, {Koster}, {Kuniyoshi}, {Kramer},
  {Kuper}, {Lambropoulos}, {Law}, {van Leeuwen}, {Lemaitre}, {Loose}, {Maat},
  {Macario}, {Markoff}, {Masters}, {McFadden}, {McKay-Bukowski}, {Meijering},
  {Meulman}, {Mevius}, {Middelberg}, {Millenaar}, {Miller-Jones}, {Mohan},
  {Mol}, {Morawietz}, {Morganti}, {Mulcahy}, {Mulder}, {Munk}, {Nieuwenhuis},
  {van Nieuwpoort}, {Noordam}, {Norden}, {Noutsos}, {Offringa}, {Olofsson},
  {Omar}, {Orr{\'u}}, {Overeem}, {Paas}, {Pandey-Pommier}, {Pandey}, {Pizzo},
  {Polatidis}, {Rafferty}, {Rawlings}, {Reich}, {de Reijer}, {Reitsma},
  {Renting}, {Riemers}, {Rol}, {Romein}, {Roosjen}, {Ruiter}, {Scaife}, {van
  der Schaaf}, {Scheers}, {Schellart}, {Schoenmakers}, {Schoonderbeek},
  {Serylak}, {Shulevski}, {Sluman}, {Smirnov}, {Sobey}, {Spreeuw}, {Steinmetz},
  {Sterks}, {Stiepel}, {Stuurwold}, {Tagger}, {Tang}, {Tasse}, {Thomas},
  {Thoudam}, {Toribio}, {van der Tol}, {Usov}, {van Veelen}, {van der Veen},
  {ter Veen}, {Verbiest}, {Vermeulen}, {Vermaas}, {Vocks}, {Vogt}, {de Vos},
  {van der Wal}, {van Weeren}, {Weggemans}, {Weltevrede}, {White}, {Wijnholds},
  {Wilhelmsson}, {Wucknitz}, {Yatawatta}, {Zarka}, {Zensus}, \& {van
  Zwieten}}]{vwg+13}
{van Haarlem}, M.~P., {Wise}, M.~W., {Gunst}, A.~W., {et~al.} 2013, \aap, 556,
  A2

\bibitem[{{Will}(1993)}]{wil93}
{Will}, C.~M. 1993, {Theory and Experiment in Gravitational Physics}

\end{thebibliography}

\clearpage

\begin{table*}
\centering
\caption{Observing Set-Ups Used\label{tab:obssetup}}
\begin{tabular}{lccccc}
    \hline
     & \multicolumn{5}{c}{Receiver} \\
    \cline{2-6}\\[-2mm]
    Parameter          & P200mm           & P217mm           & S110mm           & S60mm & S36mm \\
                       &                  & (Central beam)   &                  &       &       \\
    \hline
   Receiver band (MHz) & 1290   -- 1430   & 1240   -- 1480   & 2599.5 -- 2679.5 & 4600 -- 5100 & 7900 -- 9000 \\
System Temperature (K) & 21, 27$^{a}$ & 23 & 17 & 27 & 22 \\
Gain (K/Jy)            & 1.55  & 1.37  & 1.5   & 1.55  & 1.35 \\
   Recorded band (MHz) & 1247.5 -- 1447.5 & 1247.5 -- 1447.5 & 2527 -- 2727     & 4607.8 -- 5107.8 & 8107.8 -- 8607.8\\ %4607.812 -- 5107.812 & 8107.812 -- 8607.812 \\
Usable bandwidth (MHz) & 140              & 200              & 80               & 500 & 500 \\
   Number of sub-bands & 8                & 8                & 8                & 32    & 32 \\
    \hline
\end{tabular}
\raggedright
\\\textsc{Notes.} --- All of these receivers have circularly polarised feeds. Also, in all cases 1024 phase bins were recorded across each pulse profile.
\\$^{a}$ The reported temperatures for the two polarisation channels.\\
\end{table*}

\begin{table*}
    \caption{Summary of monthly monitoring observations of EPTA pulsars at 1.4\,GHz\label{tab:obssumm 21cm}}
\begin{tabular}{lccXcccd{3.0}cccc}
\hline
Pulsar     & Period & DM & \multicolumn{1}{c}{Obs. Span} & $N_\mathrm{obs.}$ & $N_\mathrm{det.}$ & $N_\mathrm{cal.}$ & \multicolumn{1}{c}{$\left < S/N \right >^{a}$} & \multicolumn{1}{c}{$\left < S \right >^{b}$} & \multicolumn{1}{c}{${S_\mathrm{med.}}$} & \multicolumn{1}{c}{${S_\mathrm{pub.}}^{c}$} & Ref.$^{d}$ \\ 
           &  (ms) & (\dmunit) & \multicolumn{1}{c}{(YYYY/MM)}    &  & &                     &  &\multicolumn{1}{c}{(mJy)} &  \multicolumn{1}{c}{(mJy)} & \multicolumn{1}{c}{(mJy)} & \\
\hline
J0023+0923 & 3.05 & 14.3 & 2012/03~-~2014/10 &   37 &   32 &   13 & 32 & 0.57(9) & 0.47 & \multicolumn{1}{c}{$-$} & \multicolumn{1}{c}{$-$} \\
J0030+0451 & 4.87 & 4.3  & 2011/05~-~2015/02 &   46 &   46 &   11 & 48 & 1.20(16) & 1.08 & 0.6(2) & (1) \\
J0218+4232 & 2.32 & 61.3  & 2011/04~-~2015/02 &   48 &   48 &   11 & 61 & 0.97(6) & 1.01 & 0.9(2) & (2) \\
J0340+4129 & 3.30 & 49.6  & 2011/04~-~2015/02 &   44 &   44 &   10 & 34 & 0.45(4) & 0.43 & \multicolumn{1}{c}{$-$} & \multicolumn{1}{c}{$-$} \\
J0348+0432 & 39.12 & 40.5  & 2011/07~-~2015/04 &  354 &  231 &   80 & 38 & 0.51(2) & 0.48 & \multicolumn{1}{c}{$-$} & \multicolumn{1}{c}{$-$} \\
J0613$-$0200 & 3.06 & 38.8 & 2011/03~-~2015/02 &   62 &   61 &   13 & 133 & 1.91(9) & 1.87 & 2.3$^{e}$ & (3) \\
J0621+1002 & 28.85 & 36.6 & 2011/03~-~2015/02 &   56 &   55 &   14 & 92 & 1.34(7) & 1.31 & 1.9(3) & (2) \\
J0751+1807 & 3.48 & 30.2 & 2011/01~-~2015/02 &   89 &   81 &   23 & 106 & 1.07(7) & 1.02 & 3.2(7) & (2) \\
J1012+5307 & 5.26 & 9.0 & 2011/03~-~2015/02 &   58 &   55 &   12 & 173 & 3.8(7) & 3.3 & 3(1) & (2) \\
J1022+1001 & 16.45 & 10.3 & 2011/01~-~2015/02 &  119 &  114 &   16 & 308 & 2.3(8) & 1.0 & 6.1$^{e}$ & (3) \\
J1024$-$0719 & 5.16 & 6.5 & 2011/03~-~2015/02 &   60 &   57 &   15 & 113 & 2.1(5) & 1.1 & 1.5$^{e}$ & (3) \\
J1518+4904 & 40.93 & 11.6 & 2011/03~-~2014/10 &  226 &  190 &  149 & 379 & 2.5(2) & 1.7 & 4(2) & (2) \\
J1600$-$3053 & 3.60 & 52.3 & 2011/03~-~2015/02 &   45 &   45 &    9 & 136 & 1.79(5) & 1.78 & 2.5$^{e}$ & (3) \\
J1640+2224 & 3.16 & 18.4 & 2011/03~-~2015/02 &   83 &   72 &   11 & 66 & 0.4(1) & 0.3 & 2(1) & (2) \\
J1643$-$1224 & 4.62 & 62.4 & 2011/03~-~2015/02 &   47 &   47 &   13 & 276 & 4.2(1) & 4.3 & 4.8$^{e}$ & (3) \\
J1713+0747 & 4.57 & 16.0 & 2011/03~-~2015/02 &  105 &   96 &   21 & 746 & 4.9(1.6) & 2.4 & 10.2$^{e}$ & (3) \\
J1730$-$2304 & 8.12 & 9.6 & 2011/03~-~2015/02 &   50 &   48 &   10 & 239 & 5.1(1.4) & 3.6 & 3.9$^{e}$ & (3) \\
J1738+0333 & 5.85 & 33.8 & 2011/06~-~2015/02 &   37 &   36 &    7 & 44 & 0.52(5) & 0.50 & \multicolumn{1}{c}{$-$} & \multicolumn{1}{c}{$-$} \\
J1741+1351 & 3.75 & 24.0 & 2011/10~-~2015/02 &   38 &   36 &    9 & 29 & 0.50(6) & 0.56 & 0.93$^{f}$ & (4) \\
J1744$-$1134 & 4.07 & 3.1 & 2011/03~-~2014/12 &   49 &   48 &   10 & 193 & 1.9(6) & 1.1 & 3.1$^{e}$ & (3) \\
J1853+1303 & 4.09 & 30.6 & 2013/05~-~2015/02 &   20 &   18 &    7 & 34 & 0.6(1) & 0.5 & 0.4(2) & (5) \\
B1855+09   & 5.36 & 13.3 & 2011/03~-~2015/02 &   47 &   45 &   10 & 177 & 3.6(8) & 2.3 & 5.0$^{e}$ & (3) \\
J1911+1347 & 4.63 & 31.0 & 2013/05~-~2015/02 &   22 &   21 &    7 & 57 & 0.87(15) & 0.71 & 0.08$^{f}$ & (6) \\
J1918$-$0642 & 7.65 & 26.6 & 2011/03~-~2015/02 &   45 &   45 &   10 & 67 & 1.5(3) & 1.2 & 0.58(2)$^{g}$ & (7) \\
B1937+21   & 1.56 & 71.0 & 2011/03~-~2015/02 &   84 &   75 &   11 & 524 & 12(1) & 12 & 13.2$^{e}$ & (3) \\
J2010$-$1323 & 5.22 & 22.2 & 2012/09~-~2015/02 &   25 &   24 &    8 & 70 & 0.64(7) & 0.58 & 1.6$^{f}$ & (4) \\
J2017+0603 & 2.90 & 23.9 & 2011/04~-~2015/02 &   44 &   36 &   10 & 25 & 0.48(9) & 0.40 & 0.5(2) & (8) \\
J2043+1711 & 2.38 & 20.7 & 2011/04~-~2015/02 &   44 &   19 &    7 & 15 & 0.246(16) & 0.237 & \multicolumn{1}{c}{$-$} & \multicolumn{1}{c}{$-$} \\
J2145$-$0750 & 16.05 & 9.0 & 2011/03~-~2015/02 &   52 &   50 &   10 & 382 & 2.9(5) & 2.6 & 8.9$^{e}$ & (3) \\
J2229+2643 & 2.98 & 23.0 & 2011/03~-~2015/02 &   48 &   40 &    9 & 98 & 0.5(1) & 0.5 & 0.9(2) & (2) \\
J2234+0944 & 3.63 & 17.8 & 2011/11~-~2014/10 &   42 &   39 &   15 & 62 & 0.94(14) & 0.93 & \multicolumn{1}{c}{$-$} & \multicolumn{1}{c}{$-$} \\
J2317+1439 & 3.45 & 21.9 & 2011/03~-~2015/02 &   52 &   46 &   11 & 54 & 0.8(3) & 0.3 & 4(1) & (2) \\
J2322+2057 & 4.81 & 13.4 & 2013/07~-~2015/02 &   21 &   17 &    6 & 20 & 0.5(2) & 0.3 & \multicolumn{1}{c}{$-$} & \multicolumn{1}{c}{$-$} \\
\hline
\end{tabular}
\raggedright
\\$^{a}$ Mean \snr, computed using \snr values only from observations longer than 20 minutes and scaled to a canonical integration time of 30 minutes.
\\$^{b}$ Mean phase-averaged flux density. The uncertainty reported is the standard error on the mean (i.e. $\sigma_{S_\mathrm{mean}}/\sqrt{N_\mathrm{cal.}}$). 
\\$^{c}$ Previously published phase-averaged flux density.
\\$^{d}$ References for $S_\mathrm{pub.}$ -- (1): \citet{lzb+00}, (2): \citet{kxl+98}, (3): \citet{mhb+13}, (4): \citet{jbo+07}, (5): \citet{sfl+05}, (6): \citet{lfl+06}, (7): \citet{jsb+10}, (8): \citet{cgj+11}
\\$^{e}$ \citet{mhb+13} report the RMS of multiple flux density measurements. This does not represent the uncertainty on the mean, but rather how much scintillation can modulate the observed flux density.
\\$^{f}$ No uncertainty reported.
\\$^{g}$ The flux density of PSR~J1918$-$0642 reported by \citet{jsb+10} is for a single observation. Thus, the value is likely affected by scintillation, but the uncertainty does not take scintillation into account.
\end{table*}

\begin{table*}
    \caption{Summary of monthly monitoring observations of EPTA pulsars at 2.6\,GHz \label{tab:obssumm 11cm}}
\begin{tabular}{lccXccd{3.0}}
\hline
Pulsar     & Period & DM & \multicolumn{1}{c}{Obs. Span} & $N_\mathrm{obs.}$ & $N_\mathrm{det.}$ & \multicolumn{1}{c}{$\left < S/N \right >^{a}$} \\ 
           &  (ms) & (\dmunit) & \multicolumn{1}{c}{(YYYY/MM)}    &  & &                      \\
\hline
J0023+0923 & 3.05 & 14.3 & \mathrm{2012/03}~-~\mathrm{2014/10} &   28 &    3   & 14 \\
J0030+0451 & 4.87 & 4.3 & \mathrm{2011/05}~-~\mathrm{2015/02} &   40 &   34    & 13 \\
J0218+4232 & 2.32 & 61.3 & \mathrm{2011/05}~-~\mathrm{2015/02} &   17 &    0   & -  \\
J0340+4129 & 3.30 & 49.6 & \mathrm{2011/05}~-~\mathrm{2015/02} &   36 &   13   & 10 \\
J0613$-$0200 & 3.06 & 38.8 & \mathrm{2011/03}~-~\mathrm{2015/02} &   42 &   37 & 20 \\
J0621+1002 & 28.85 & 36.6 & \mathrm{2011/03}~-~\mathrm{2015/02} &   39 &   29  & 15 \\
J0751+1807 & 3.48 & 30.2 & \mathrm{2011/03}~-~\mathrm{2015/02} &   49 &   35   & 30 \\
J1012+5307 & 5.26 & 9.0 & \mathrm{2011/03}~-~\mathrm{2015/02} &   48 &   46    & 39 \\
J1022+1001 & 16.45 & 10.3 & \mathrm{2011/03}~-~\mathrm{2015/02} &   60 &   54  & 109 \\
J1024$-$0719 & 5.16 & 6.5 & \mathrm{2011/05}~-~\mathrm{2015/02} &   46 &   39  & 20 \\
J1518+4904 & 40.93 & 11.6 & \mathrm{2011/03}~-~\mathrm{2014/06} &    7 &    6  & 63 \\
J1600$-$3053 & 3.60 & 52.3 & \mathrm{2011/05}~-~\mathrm{2014/12} &   31 &   29 & 25 \\
J1640+2224 & 3.16 & 18.4 & \mathrm{2011/06}~-~\mathrm{2015/02} &   48 &   32   & 13 \\
J1643$-$1224 & 4.62 & 62.4 & \mathrm{2011/05}~-~\mathrm{2015/01} &   36 &   36 & 57 \\
J1713+0747 & 4.57 & 16.0 & \mathrm{2011/05}~-~\mathrm{2015/02} &   46 &   44   & 258 \\
J1730$-$2304 & 8.12 & 9.6 & \mathrm{2011/05}~-~\mathrm{2015/02} &   35 &   24  & 48 \\
J1738+0333 & 5.85 & 33.8 & \mathrm{2011/06}~-~\mathrm{2015/01} &    9 &    2   & 32 \\
J1741+1351 & 3.75 & 24.0 & \mathrm{2012/01}~-~\mathrm{2015/02} &   35 &   15   & 17 \\
J1744$-$1134 & 4.07 & 3.1 & \mathrm{2011/05}~-~\mathrm{2015/01} &   39 &   37  & 27 \\
J1853+1303 & 4.09 & 30.6 & \mathrm{2013/05}~-~\mathrm{2015/02} &   24 &    6   & 10 \\
B1855+09   & 5.36 & 13.3 & \mathrm{2011/03}~-~\mathrm{2015/02} &   43 &   35   & 46 \\
J1911+1347 & 4.63 & 31.0 & \mathrm{2013/05}~-~\mathrm{2015/02} &   22 &   17   & 14 \\
J1918$-$0642 & 7.65 & 26.6 & \mathrm{2011/05}~-~\mathrm{2015/02} &   41 &   24 & 22 \\
B1937+21   & 1.56 & 71.0 & \mathrm{2011/03}~-~\mathrm{2015/02} &   54 &   48   & 87 \\
J2010$-$1323 & 5.22 & 22.2 & \mathrm{2012/09}~-~\mathrm{2015/02} &   26 &   18 & 17 \\
J2017+0603 & 2.90 & 23.9 & \mathrm{2011/07}~-~\mathrm{2015/02} &   16 &    1   & 11 \\
J2043+1711 & 2.38 & 20.7 & \mathrm{2011/05}~-~\mathrm{2015/02} &   17 &    1   & 6 \\
J2145$-$0750 & 16.05 & 9.0 & \mathrm{2011/03}~-~\mathrm{2015/02} &   47 &   45 & 70 \\
J2229+2643 & 2.98 & 23.0 & \mathrm{2011/03}~-~\mathrm{2015/02} &   46 &   35   & 25 \\
J2234+0944 & 3.63 & 17.8 & \mathrm{2011/11}~-~\mathrm{2014/10} &   36 &   28   & 21 \\
J2317+1439 & 3.45 & 21.9 & \mathrm{2011/03}~-~\mathrm{2015/02} &   43 &   28   & 14 \\
J2322+2057 & 4.81 & 13.4 & \mathrm{2013/07}~-~\mathrm{2015/02} &   16 &    3   & 8 \\
\hline
\end{tabular}
\raggedright
\\$^{a}$ Mean \snr, computed using \snr values only from observations longer than 20 minutes and scaled to a canonical integration time of 30 minutes.
\end{table*}

\begin{table*}
\centering
\caption{High-frequency observations\label{tab:highfreq}}
\begin{tabular}{llcd{2.2}d{1.3}d{2.4}}
    \hline
    Pulsar       & \hdr{Obs. Start} & Integration Time & \hdr{\snr} & \hdr{Flux Density} & \hdr{TOA Uncertainty} \\
                 & \hdr{(UTC)} &       (s)   &      &   \hdr{(mJy)}     &      \hdr{($\mu$s)}   \\
    \hline
    \multicolumn{6}{c}{\textit{5-GHz Observations}} \\
    \hline
    J0751+1807   & 2015-Jan-28 02:12:19 & 2960 &  47.05 & 0.375 & 2.890 \\
    J1012+5307   & 2015-Jan-24 20:03:30$^{a}$ & 3740 & 15.96 & 0.209 & 7.278 \\
                 & 2015-Jan-28 03:11:00 & 2060 &  26.22 & 0.385 & 4.866 \\
    J1022+1001   & 2015-Jan-24 21:01:09 & 1780 &  39.47 & 0.562 & 3.552 \\
    J1518+4904   & 2015-Jan-28 03:59:29 & 1770 &  25.51 & 0.441 & 5.87  \\
    J1600$-$3053 & 2015-Jan-28 06:38:40 & 1770 &  10.49 & 0.318 & 6.657 \\
    J1643$-$1224 & 2015-Jan-26 07:14:40 & 1780 &  22.78 & 0.343 & 4.920 \\
    J1713+0747   & 2015-Jan-26 06:13:59 & 1200 & 133.87 & 1.463 & 0.470 \\
                 & 2015-Jan-28 04:38:09 & 1780 & 110.45 & 1.102 & 0.541 \\
    J1730$-$2304 & 2015-Jan-26 07:51:49 & 1770 &  27.10 & 0.503 & 5.892 \\
    J1744$-$1134 & 2015-Jan-28 06:09:40 & 1280 &  14.40 & 0.389 & 1.421 \\
    B1855+09     & 2015-Jan-26 08:29:29 & 1780 &  52.50 & 0.895 & 1.482 \\
    B1937+21     & 2015-Jan-07 15:13:19 & 2310 &  86.56 & 1.234 & 0.101 \\
                 & 2015-Jan-28 05:26:40 & 1780 &  35.91 & 0.639 & 0.211 \\
    J2145$-$0750 & 2015-Jan-07 15:59:59 & 1780 &  52.70 & 0.702 & 3.812 \\
                 & 2015-Jan-25 15:20:19 & 1780 &  54.00 & 0.797 & 3.883 \\
    \hline
    \multicolumn{6}{c}{\textit{9-GHz observations}} \\
    \hline
    J1022+1001   & 2015-Jan-24 22:09:30 & 2610 &  10.06 & 0.233 & 18.237 \\
    J1713+0747   & 2015-Jan-26 06:38:40 & 1780 &  15.74 & 0.258 & 5.297 \\
    J2145$-$0750 & 2015-Jan-07 17:11:10 & 2950 &  16.90 & 0.251 & 12.269 \\
                 & 2015-Jan-26 13:37:39 & 2680 &  16.43 & 0.218 & 10.631 \\
    \hline
\end{tabular}
\raggedright
\\$^{a}$ Two closely spaced observations added together.\\
\end{table*}

\begin{table*}
\centering
\caption{Parameters for selected available and planned observing set-ups\label{tab:otherrcvr}}
\begin{tabular}{lcccc}
    \hline
                       & \multicolumn{4}{c}{Observing Set-Up} \\
    \cline{2-5} \\[-2mm]
    Telescope          & Effelsberg         & Effelsberg     & MeerKAT           & LOFAR \\
    Receiver           & UBB                & S45mm (``C+'') & S-band            & DE601$^{a}$ \\
   Receiver band (MHz) & \phantom{1}600 -- 2700$^{b}$ & 4000 -- 9300   & 1600 -- 3500      & 111.5 -- 186.5 \\
   Recorded band (MHz) & 1100 -- 2700\phantom{$^{b}$}       & 4000 -- 6000   & 1600 -- 2300 & 111.5 -- 186.5 \\
System Temperature (K) & 25/(55)$^{c}$      & $\sim$25$^{d}$   & $<$25$^{e}$       & \phantom{$^{f}$}--$^{f}$ \\
Gain (K/Jy)            & 1.25               & $\sim$1.35$^{d}$ & 2.5          & \phantom{$^{f}$}--$^{f}$ \\
SEFD (Jy)              & 45                 & 18.5           & 10                & \phantom{$^{f}$}1500$^{f}$ \\
    \hline\\
\end{tabular}
\raggedright
\\$^{a}$ DE601 is the international LOFAR station located at Effelsberg.
\\$^{b}$ The observing band of the UBB receiver currently goes as low as
600\,MHz. However, strong interference from digital television broadcasts at
$\sim$500 to 800\,MHz and GSM emission at $\sim$900\,MHz greatly deteriorates
the quality of the observations. There are plans to insert a high-pass filter
at $\sim1100$\,MHz.  We use this planned low-frequency band for our estimates. 
\\$^{c}$ The strong interference in the UBB band causes the system temperature
to be on the order of $\sim$55\,K, significantly higher than the value
of $T_\mathrm{sys} = 25$\,K measured in the lab. A high-pass
filter will be installed around 1100\,MHz and should lower $T_\mathrm{sys}$ to
$\sim25$\,K. Nevertheless, in our estimates of the system performance, we use
the current, higher, measured value to be conservative.
\\$^{d}$ Preliminary estimate.
\\$^{e}$ This is the target system temperature. In our estimates we use
$T_\mathrm{sys} = 25$\,K.
\\$^{f}$ We estimated the SEFD Effelsberg LOFAR station by scaling the SEFD of
the LOFAR core published by \citet{vwg+13}.
\end{table*}

% Table data for AO C-band receiver
% Arecibo  
% C-band   
% 3950 -- 6050 
% 3950 -- 4750 
% $\sim$28 
% $\sim8$  
% 3.5      

\end{document}